\documentclass[11pt, a4paper]{article}

\usepackage{color}
\usepackage{graphicx}
\usepackage{amssymb}
\usepackage{amsmath}
\usepackage[english]{babel}
\usepackage[T1]{fontenc}
\usepackage[noadjust]{cite}
\usepackage{hyperref} 

\usepackage[utf8]{inputenc}
\usepackage{authblk}

\newcommand\eeq{\end{equation}}
\newcommand\beq{\begin{equation}}
\newcommand\eea{\end{eqnarray}}
\newcommand\bea{\begin{eqnarray}}

\begin{document}

\linespread{1.1}

\title{ \color{red} \bf Dark matter and dark radiation\\ from evaporating primordial black holes }

\author[1,2]{ {\Large Isabella Masina} \thanks{masina@fe.infn.it}}

\affil[1]{\small Dip. di Fisica e Scienze della Terra, Ferrara University and INFN, Ferrara, Italy }
\affil[2]{Theoretical Physics Department, CERN, Geneva, Switzerland}
\date{}

\maketitle

\begin{abstract}
Primordial black holes in the mass range from $10^{-5}$ to $10^9$ g 
 might have existed in the early universe. Via their evaporation mechanism (completed before Big Bang Nucleosynthesis), they might have released stable particles beyond the Standard Model. We reconsider the possibility that such particles might constitute the main part or a fraction of the dark matter observed today, updating the impact on this scenario from warm dark matter constraints. If sufficiently light, stable particles from primordial black holes evaporation might also provide a significant contribution to dark radiation. We generalize previous studies on this interesting dark matter and dark radiation production mechanism, by including the effects of accretion and a possible amount of entropy non conservation. We also discuss in some detail specific examples of stable particle candidates beyond the Standard Model.

\end{abstract}

\linespread{1.2}

\vskip 1.cm
\section{Introduction}

Primordial black holes (PBHs) with masses in the broad range $10^{-5}-10^9$ g might have existed in the early universe. 
Thought they completely evaporated before Big Bang Nucleosynthesis (BBN), their past existence might have had a deep impact on the  cosmological dynamics because of their mechanism of evaporation\,\cite{Hawking:1974sw}, according to which all particle states with masses below the Hawking temperature are produced.
Apart from the case of gravitino production \cite{Khlopov:2004tn, Khlopov:2008qy}, for this range of masses, the PBHs density at formation is at present unconstrained, as reviewed e.g. in ref.\,\cite{Carr:2020gox}.
A method to constrain the range $10^3-10^5$ g has been recently presented\,\cite{Inomata:2020lmk}. 

The proposal of the early existence of collapsed objects, later called PBHs, dates back to 1967\,\cite{ZelNov:1967}. 
The formation of PBHs from early universe inhomogeneities was considered in refs.\,\cite{Hawking:1971ei, Carr:1974nx, Hawking:1974sw}. However, since inflation removes all pre-existing inhomogeneities, any cosmologically interesting PBH density 
has to be created after inflation. Various mechanisms have been proposed, as for instance:
that they formed from large inhomogeneities arising through quantum effects during inflation; 
that some sort of phase transition may have enhanced PBHs formation from primordial inhomogeneities or triggered it. 
We refer to\,\cite{Carr:2020gox, Khlopov:2013ava} for reviews of these proposals, with proper references to the associated literature.

Once formed, PBHs start the process of evaporation. The particles produced might be responsible for the excess of baryons over anti-baryons\,\cite{Hawking:1974rv, Zeldovich:1976vw}. 
Among the products of PBHs evaporation, there might also be (cosmologically) stable particles beyond the Standard Model (SM): 
such particles would contribute to the observed dark matter (DM) abundance\,\cite{Fujita:2014hha, Lennon:2017tqq, Morrison:2018xla}
and, if sufficiently light, also to dark radiation (DR) \,\cite{Lennon:2017tqq, Hooper:2019gtx, Lunardini:2019zob, Hooper:2020evu}. 

The aim of this paper is to provide a complete and updated study of the possible contribution to DM and DR by cosmologically stable  particles beyond the SM produced in the process of PBHs evaporation, considering both scenarios of radiation and BH domination (more on this later). 
We explicitly include the effect of accretion\footnote{Accretion was considered in\,\cite{Hooper:2019gtx} 
in an implicit way: the BH mass parameter used there refers to the mass of a given BH after accretion and merging have ceased to be efficient.},
update the constraints from warm DM and discuss ways to overcome them, as for instance a possible non conservation of entropy, as suggested in\, \cite{Fujita:2014hha}. 
We also discuss in some detail the connection with specific beyond SM scenarios. 
The literature on DM and DR from PBHs, also in connection with baryogenesis, is quite rich, but given the present interest in this field,
we think that a detailed and updated study can be useful.

The possibility of an early universe epoch during which the energy content of the universe was dominated by PBHs, called BH domination for short, was suggested by Barrow et al.\,\cite{Barrow:1990he} studying baryogenesis from PBHs. The scenario of BH domination is interesting because the final asymmetry is independent on the initial PBHs number density. Further studies extended this scenario by considering Planck scale relics as DM\,\cite{Baumann:2007yr} and leptogenesis from the evaporation of PBHs\,\cite{Baumann:2007yr, Fujita:2014hha}. 

In particular, Fujita et al.\,\cite{Fujita:2014hha}, assuming PBHs domination, calculated the contribution to DM by new particles beyond the SM: they found that a significant contribution to DM could come from stable particles that are either superheavy or with masses in the MeV range; the latter light DM candidates would be warm, and the lower limits on their mass coming from the warm DM velocity constraints\,\cite{Viel:2005qj} were also discussed in ref.\,\cite{Fujita:2014hha}.  
A more sophisticated study of such lower limits was done by Lennon et al.\,\cite{Lennon:2017tqq}, confirming the order of magnitude results of ref.\,\cite{Fujita:2014hha}. 
Focussing on the radiation domination scenario, Morrison et al.\,\cite{Morrison:2018xla} also studied DM from PBHs,  
in relation with baryogenesis and leptogenesis.

Hooper et al.\,\cite{Hooper:2019gtx} recently pointed out that, if there was an epoch dominated by PBHs, such particles might significantly contribute to the DR\footnote{This also allows to alleviate the tension with $H_0$ measurements. For an approach based on the SM see instead\,\cite{Nesseris:2019fwr}.} (this work also considered the superheavy DM case, but not the light one).
Lunardini et al.\,\cite{Lunardini:2019zob} also studied DR from PBHs evaporation, focussing on light neutrinos with Dirac or Majorana nature. The effect of PBHs merging was recently reconsidered in ref.\,\cite{Hooper:2020evu}, showing that a significant quantity of high-energy gravitons might be produced by such mechanism.

In this paper we want to reconsider in a complete and updated way the possibility that evaporating PBHs might provide a significant fraction of the DM observed today and might also contribute to DR. 
The paper is organized as follows. In sec. 2 we introduce our notation and review basic ideas about formation of PBHs. In sec. 3
we discuss the mechanisms of accretion and evaporation. In sec. 4 we calculate the lifetime of the PBHs, and discuss the dynamics
of their early abundance in sec. 5. Sec. 6 deals with the characteristics of the particles produced in the evaporation of PBHs, updating the bounds on warm DM. The calculation of the contribution to DM and DR from stable particles emitted by PBHs is presented in secs. 7 and 8 respectively. A discussion of the results and the conclusions are presented in sec. 9.

We do not use natural units, in order to have formulas more ready to use for numerical computations.


\section{PBHs in the early Universe}

\subsection{Preliminaries and radiation dominated era }

According to the first Friedmann equation, neglecting the curvature and cosmological constant terms, the early universe evolution is described by
\beq
 \left( \frac{\dot a}{a} \right)^2   \equiv H(t)^2 =\frac{8 \pi G}{3 }   \rho(t)\,,
\label{eq-F1}
\eeq
where $a(t)$ is the scale factor, $H(t)$ is the Hubble parameter, $\rho(t)$ is the mass density of the Universe 
and  $G$ is the Newton gravitational constant, $G \simeq 6.674 \times 10^{-11} \,\rm{ m^3 / (kg \,s^2)}$.

In the early hot and dense universe, it is appropriate to assume an equation of state corresponding to a gas of radiation (or relativistic particles). 
During radiation domination, $\rho \propto a^{-4}$, $a(t) \propto t^{1/2}$, and
\beq
H(t)= \frac{1}{2 t }\,\,.
\label{eq-raddom}
\eeq

At relatively late times, non-relativistic matter eventually dominates the mass density over radiation. 
A pressureless gas 
leads to the expected dependence $\rho \propto a^{-3}$, $a(t) \propto t^{2/3}$, and
\beq
H(t) = \frac{2}{3 t} \,.
\eeq

In general, the radiation mass density (at high temperatures) can be approximated by including only those particles which are in thermal equilibrium and have masses below the temperature of the radiation bath:
\beq
 \rho_R = \frac{\pi^2 g_*(T)  } {30}  \frac{(k_B T )^4 }{(\hbar \,c)^3 \,c^2} \,\,\, , \,\,\, \,\,
 g_*(T)= \sum_B g_B + \frac{7}{8}\sum_F g_F \,\,\, , 
 \label{eq-rhoRT}
\eeq
where $k_B$ is the Boltzmann constant, $k_B \simeq 8.617 \times 10^{-5} $ eV/K, 
$\hbar$ is the reduced Planck constant, $\hbar=6.582 \times 10^{-16}$ eV s, 
$c$ is the velocity of light in vacuum, $c=2.998 \times 10^{8}$ m/s,
and $g_{B(F)}$ is the number of degrees of freedom (dofs) of each boson (fermion). 
Below the electron mass,
$g_*(T)=7.25$. 
For the full SM\footnote{Adding to the SM three light right-handed neutrinos (as in the case of neutrinos with Dirac nature or in the case of a low-scale seesaw mechanism),  $g_*(T)= 112$.
At higher temperatures, $g_*(T)$ will be model-dependent. 
For example, in the standard seesaw mechanism, $g_*(T)= 112$ above the scale corresponding to the Majorana mass of the three heavy right-handed neutrinos.
In the minimal $SU(5)$ model, $g_*(T) = 160.75$ at temperatures above the GUT scale; including also three heavy right-handed neutrinos $g_*(T) = 166$. In a supersymmetric model, at temperatures above the SUSY mass scale,
$g_{*}(T)$ would at least double with respect to the non-supersymmetric case.},
here defined including three light left-handed neutrinos, $g_*(T)= 106.75$.

Assuming radiation domination, the relation between temperature and time is
\beq
 k_B T   =   \left( \frac{45} {16 \pi^3 g_*(T)  } \right)^{1/4}  (M_{Pl} c^2)^{1/2} \left( \frac{\hbar}{ t} \right)^{1/2} 
 \approx 5\times 10^{13}\, {\rm GeV} \left( \frac{10^{-34}\, {\rm s}}{t } \right)^{1/2}  \left( \frac{106.75}{g_*(T) } \right)^{1/4}  \, ,
 \label{eq-R-Tt}
\eeq
where we introduced the Planck mass, $M_{Pl}= \sqrt{ \hbar c / G } \approx 1.221 \times 10^{19}\, \rm{GeV}/c^2  \approx  2.176 \times 10^{-8} $ kg. Due to the mild dependence on  $g_*(T)$ in eq.\,(\ref{eq-R-Tt}), the relation between temperature and time is not significantly modified  in models with additional dofs with respect to the SM.

\subsection{Formation of PBHs}

As reviewed for instance in ref. \cite{Carr:2020gox}, if a PBH forms in the radiation dominated era,
typically its mass is close to the value enclosed by the particle horizon near the end of inflation:
\beq
M_{BH} =  \gamma   \frac{4 \pi}{3}  \rho \left( 2\, c \,t_f \right)^3 =\gamma   \frac{4 \pi}{3}  \rho  \left( \frac{ c }{H_f}   \right)^3\,,
\label{eq-MPBH}
\eeq
where $\gamma \lesssim 1$ is a numerical factor that depends on the details of the gravitational collapse, $\rho$ is the radiation density, $t_f$ and $H_f$ are respectively the cosmic time and the Hubble parameter at the formation of the PBH, and in the last equality we used eq.\,(\ref{eq-raddom}) assuming radiation domination. 
Using eq.\,(\ref{eq-F1}), we can also write 
\beq
 M_{BH} =   \frac{\gamma}{2} \frac{(M_{Pl}c^2) ^2}{ \hbar \,H_f}\frac{1}{c^2} \approx  \gamma   \frac{10^{10} \,\rm{GeV}} {\hbar \,H_f}  10^4 \,\rm{g}  \gtrsim \frac{\gamma}{3} \,{\rm  g} \, ,
\label{eq-PBH}
\eeq
where the last lower bound follows from the fact that CMB observations put an upper bound on the Hubble scale during inflation, $\hbar H_I \lesssim 3 \times 10^{14}$ GeV at $95\%$ C.L. \cite{Akrami:2018odb}, and we have $H_f \lesssim H_I$.
In the literature the value $\gamma=1/(3\sqrt{3})\approx 0.2$ is usually taken as reference value\,\cite{Carr:2020gox}; in this case the lower limit would become $M_{BH} \gtrsim 0.07$ g. In any case, PBHs should have mass larger than the Planck mass, namely $M_{BH}\gtrsim 10^{-5}$\,g.
As is well known (and will be reviewed in the following), there is also an upper bound on $M_{BH}$ coming from constraints on BBN:  $M_{BH} \lesssim 10^9$\,g. 
The range of PBH masses between these bounds is at present generically unconstrained \cite{Carr:2020gox}. 

\begin{figure}[t!]
\vskip .0 cm 
 \begin{center}\includegraphics[width=11 cm]{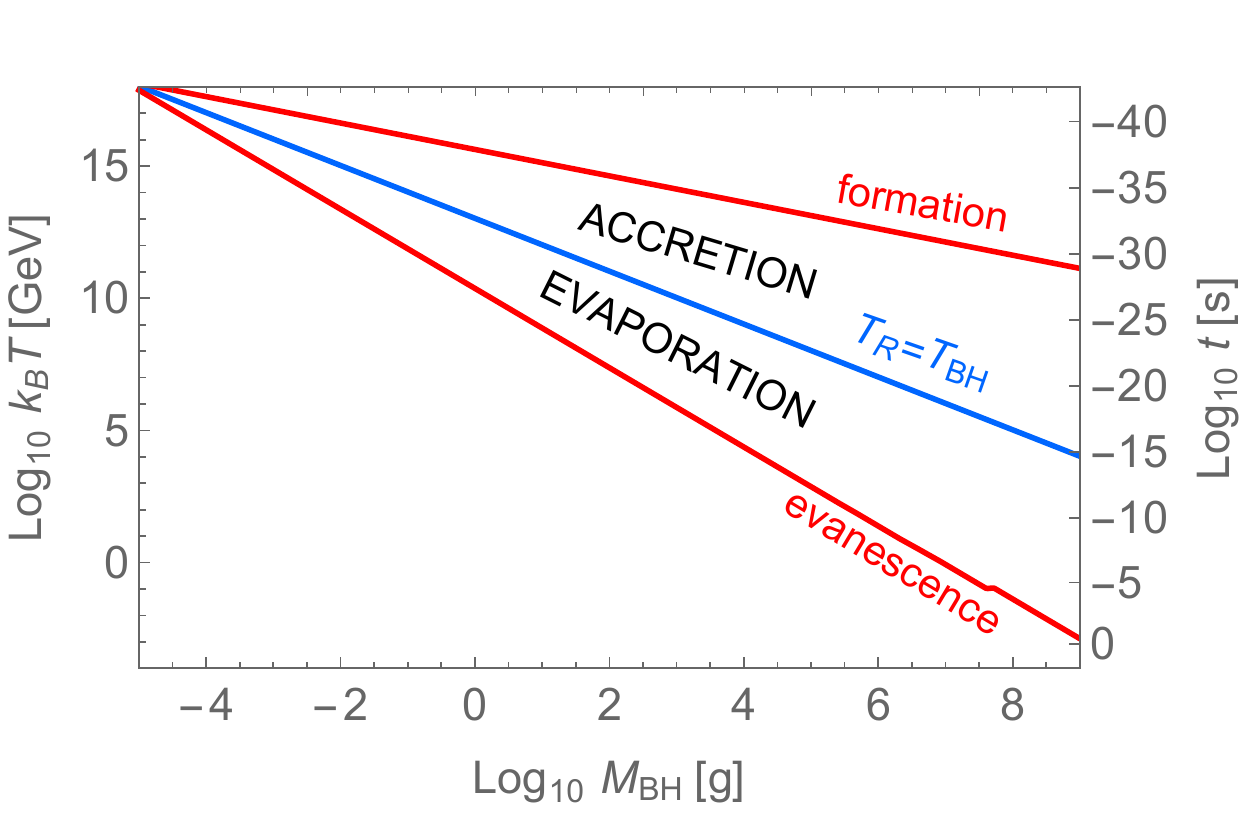}   \,\, \,
 \end{center}
\caption{\baselineskip=15 pt \small 
From top to bottom and as a function of the PBH mass at formation, $M_{BH}$, we show the formation temperature, the BH (or Hawking) temperature and, assuming radiation domination, the temperature at evanescence. 
The right vertical axis shows the relation with the cosmic time assuming radiation domination. We took $\gamma=0.2$ and the SM dofs, $g_*(T)= 106.75$. The region with $M_{BH} \gtrsim 10^9$\,g is severely constrained by BBN. The region with $M_{BH} \lesssim 0.1$ g is excluded by CMB constraints on inflation if $\gamma=0.2$; in general, values down to $M_{BH}\approx  \gamma/3$\,g are allowed (in any case, $M_{BH}\gtrsim 10^{-5}$\,g).}   
\label{fig-TBHM}
\vskip .2 cm
\end{figure}

Recalling eq.\,(\ref{eq-raddom}), the PBHs formation time is easily calculated from eq.\,(\ref{eq-PBH}):
\beq
\frac{t_{f}}{\hbar}=\frac{1} {\gamma}  \frac{M_{BH} c^2}{ (M_{Pl} c^2)^2  } \,.
\label{eq-tfBH}
\eeq
Notice that the ratio $M_{BH}/t_{f}$ is independent on $M_{BH}$.
As for the temperature at formation,
combining eqs.\,(\ref{eq-rhoRT}), (\ref{eq-F1}) and (\ref{eq-PBH}), we have 
\beq
k_B T_{f}  
 = \left(  \frac{45 \gamma^2  } {16 \pi^3 g_*(T_{f}) }   \right)^{1/4}  \left( \frac{M_{Pl}  }{M_{BH}} \right)^{1/2}
 M_{Pl} c^2 \,.
 \label{eq-Tf}
\eeq
The temperature and the time at formation of PBHs are plotted in fig.\,\ref{fig-TBHM} as a function of the PBH mass. In this and in the following plots, we take for definiteness $\gamma=0.2$ and the SM dofs, $g_*(T)= 106.75$. In this case, a $1\, (10^8)$ g PBH forms when the radiation bath has a temperature of about $10^{16} \,(10^{12})$ GeV, which corresponds to $t_f$ about $10^{-38} \, (10^{-30})$ s.

Assuming adiabatic cosmic expansion after PBH formation, the ratio of the PBH number density to the entropy
density, $n_{BH}/s$, is conserved. It is useful to introduce the parameter $\beta$ defined as
\beq
\beta= \frac{\rho_{BH}(t_f)}{\rho_R(t_f)} \, . 
\eeq

In the following, we consider the evolution of non-rotating (Schwarzschild) uncharged PBHs, see e.g. ref.\,\cite{Dai:2009hx} for a discussion of the effects related to hidden charges.

\section{Accretion vs evaporation}

Once formed, PBHs start to evaporate via the process of Hawking radiation \cite{Hawking:1974sw}. Despite this, they can even gain a mass larger than the initial one as they can: i) form binaries and merge; ii) accrete mass from the surrounding radiation bath. The first phenomenon is not very efficient\,\cite{Hooper:2019gtx}, but see ref.\,\cite{Hooper:2020evu} for a revisitation. 
We now turn to discuss  in some detail the second phenomenon, included implicitly in the analysis of\,\cite{Hooper:2019gtx} (the BH mass parameter used there does not refer to the BH mass at formation, but to the mass of a given BH after accretion and merging have ceased to be efficient).

\subsection{Accretion}

A BH in a radiation bath gains mass at a rate calculated by Bondi \cite{Bondi:1952ni}. 
As discussed e.g. in \cite{Nayak:2009wk}, it is customary to take the accretion rate to be proportional to the product of the surface area of the PBH and the mass density of radiation,
\beq
\frac{dM_{BH}}{dt} = f_{acc} \, c\,(4 \pi r_{BH}^2) \,  \rho_R \, ,
\eeq
where $f_{acc}$ is the accretion efficiency, $r_{BH}= 2 G M_{BH}/c^2$ is the Schwarzschild radius of the PBH and $\rho_R$
is the mass density of radiation. The value of $f_{acc}$ depends upon complex physical processes such as the mean free paths of the particles comprising the radiation surrounding the PBHs. 
Any peculiar velocity of the PBH with respect to the cosmic frame could increase the value of $f_{acc}$
\cite{Majumdar:1995yr, Custodio:1998pv, Custodio:1999vz, Custodio:2002gj, Mahapatra:2013bpa}. 
Since the precise value of $f_{acc}$ is unknown, it is customary to take $f_{acc}=O(1)$.

Using the expression for $\rho_R$ in eq.\,(\ref{eq-rhoRT}), we have
\beq
\frac{dM_{BH}}{dt} =   \frac{8 \pi^3  }{15} f_{acc}\,  g_*(T)\, \frac{M_{BH}^2 }{c^5\,  \hbar c M_{Pl}^4} (k_B T)^4\,  .
\label{eq-acc}
\eeq
It might also be useful to render explicit the time dependence by using eqs.\,(\ref{eq-F1}) and (\ref{eq-raddom})
\beq
\frac{dM_{BH}}{dt} =  f_{acc} \frac{3 }{2} \frac{\hbar c M_{BH}^2}{c^3 M_{Pl}^2} \frac{1}{t^2}  
\,.
\label{eq-acctime}
\eeq

\subsection{Evaporation}

PBHs evaporate by producing all particle states with masses below the Hawking temperature\,\cite{Hawking:1974sw}: 
\beq
k_B \,T_{BH} 
=\frac{ 1}{8 \pi } \frac{  (M_{Pl} c^2)^2}{ M_{BH} c^2}\,\,.
\eeq
For instance, as can be seen from fig.\,\ref{fig-TBHM}, all SM particles are produced for $M_{BH} \lesssim 10^8$ g, whereas to produce particles as heavy as $10^{10}$ GeV one needs $M_{BH} \lesssim 10^3$ g.
This means that each BH at some stage of its life, will radiate heavy particles beyond the SM, if they exist (as for instance the heavy right-handed neutrinos of the seesaw mechanism, GUT particles, supersymmetric particles, etc). If the radiated heavy particles are coupled to the SM sector, 
they will decay soon; if on the contrary they (or some of their decay products) are stable, they contribute to the DM of the universe and, if sufficiently light, also to the DR \cite{Lennon:2017tqq, Hooper:2019gtx}.

A BH looses mass via the Hawking evaporation process \cite{Hawking:1974sw} at a rate given by (see for instance \cite{Carr:2009jm, Carr:2020gox})
\beq
\frac{dM_{BH}}{dt} = -f_{ev} \, c\,(4 \pi r_{BH}^2) \,  \rho_{BH} \, ,
\eeq
where $f_{ev}$ is an efficiency factor for evaporation and $\rho_{BH}$ is defined as
\beq
\rho_{BH} = \frac{\pi^2}{30} \, \frac{ g_{BH}(T_{BH})}{4} \,  \frac{(k_B\,T_{BH} )^4 }{(\hbar \,c)^3 \,c^2} \,,
\eeq
with $g_{BH} (T_{BH})= \sum_{B,F}  g_{B,BH} + \frac{7}{8} g_{F,BH} $ counting the bosonic and fermionic dofs below $T_{BH}$.

Substituting the expression for the Schwarzschild radius and $\rho_{BH}$, we have an expression whose structure is similar to the one obtained for accretion in eq. (\ref{eq-acc}):
\beq
\frac{dM_{BH}}{dt} = -  \frac{8 \pi^3   }{15} \,\frac{f_{ev}\, g_{BH}(T_{BH})}{4}  \frac{M_{BH}^2 }{c^5\,  \hbar c M_{Pl}^4} (k_B T_{BH})^4\,  .
\label{eq-ev}
\eeq

\subsection{Mass gain due to accretion}

Comparing eqs.\,(\ref{eq-acc}) and (\ref{eq-ev}), it is clear that accretion dominates over evaporation when the temperature of the radiation bath is bigger than $T_{BH}$, $T > T_{BH}$, and is negligible when $T<T_{BH}$, as shown in fig.\,\ref{fig-TBHM}. The more the PBHs are heavy, the more the gap between $T_{f}$ and $T_{BH}$ is large. 
In particular, for $M_{BH} \sim 10^8$ g, the accretion phase stops when the temperature goes below about $10^5$ GeV. 

Since accretion dominates the BH dynamics from the formation time $t_f$ and until the time $t_{acc}$, when the radiation temperature equals the BH temperature, one can calculate the mass gain during this epoch by 
integrating eq.\,(\ref{eq-acctime}) and using eq.\,(\ref{eq-tfBH})
\beq
 R_{acc}   \equiv \frac{ M^{acc}_{BH}} { M_{BH}}
\approx \frac{1}{1 - \frac{3}{2} f_{acc} \gamma   }\,,
 \eeq
where $M^{acc}_{BH}$ is the mass of the BH at $t_{acc}$ and the approximation holds since $t_{acc} >> t_f$. 
Notice that the mass gain thus depends only on the combination $f_{acc} \gamma$. To be quantitative, taking 
$\gamma=0.2$ and $f_{acc}=(0.5,1,2,3)$, one obtains $R_{acc}=(1.17,1.40,2.36,7.46)$ respectively.

\section{PBHs lifetime}

It is useful to substitute the expression for the Hawking temperature in eq.\,(\ref{eq-ev}) and redefine the product $f_{ev} \,g_{BH}(T_{BH}) = \mathcal{G}\, g_{\star, H} (T_{BH})$, obtaining \cite{MacGibbon:1991tj}
\beq
\frac{d M_{BH}}{dt} = - \,  \frac{c^3}{\hbar c}\frac{ \mathcal{G}\, g_{\star, H} (T_{BH}) }{30720 \,\pi }
    \frac{M_{Pl}^4}{M_{BH}^2}   \,,
 \label{eq-evf}
\eeq
where ${\mathcal G} \approx 3.8$ is a graybody factor and $g_{\star, H} (T_{BH})$ counts all existing particle states with mass below $k_B T_{BH}$  according to \cite{MacGibbon:1991tj}
\beq
g_{\star, H} (T_{BH})  = \sum_{i} w_i  \, g_{i,H}  \, e^{-\frac{M_{BH}}{\beta_i M_i}}\,,
\eeq
where $w_i  =2 s_i+1$ for massive particles of spin $s_i$, $w_i=2$ for massless particles with $s_i>0$, $w_i=1$ for particles with $s_i=0$, and
\beq
g_{i,H} = 1.82, 1 \,({\rm{q=0} }) \,{\rm or}\, 0.97\, ({\rm{q=\pm e}}), 0.41, 0.14, 0.05  \,\,\,\, {\rm{for}} \,\,\,\, s_i=0,1/2,1,3/2,2
\eeq 
respectively, 
\beq
\beta_s = 2.66, 4.53, 6.04, 9.56  \,\,\,\, {\rm{for}} \,\,\,\, s_i=0,1/2,1,2\, ,
\eeq
and
$ M_i $ is the mass of a hole whose Hawking temperature equals the rest mass of the i-th species:
$ \frac{\hbar c \,c^2} {8 \pi G M_{i}}= m_i c^2$.

At BH temperatures below the MeV scale, corresponding to $M_{BH} >> 10^{16}$ g, only the massless photons and three left-handed neutrinos are emitted in the SM\footnote{If neutrinos are massive, for very high BH masses, they are not emitted.}. 
For the massless photon $g_{\gamma,H} =
0.82$, while for three massless (or relativistic) left-handed neutrinos $g_{3 \nu_L, H}= 
6$: the sum is $g_{*,H}=6.82 \sim 7$.
Adding to the SM also three right-handed neutrinos - to form a Dirac mass or a seesaw at low energy - one would have to add also $g_{3 \nu_R, H}= 6$.

\begin{figure}[h!]
\vskip .0 cm 
 \begin{center}
\includegraphics[width=7.6cm]{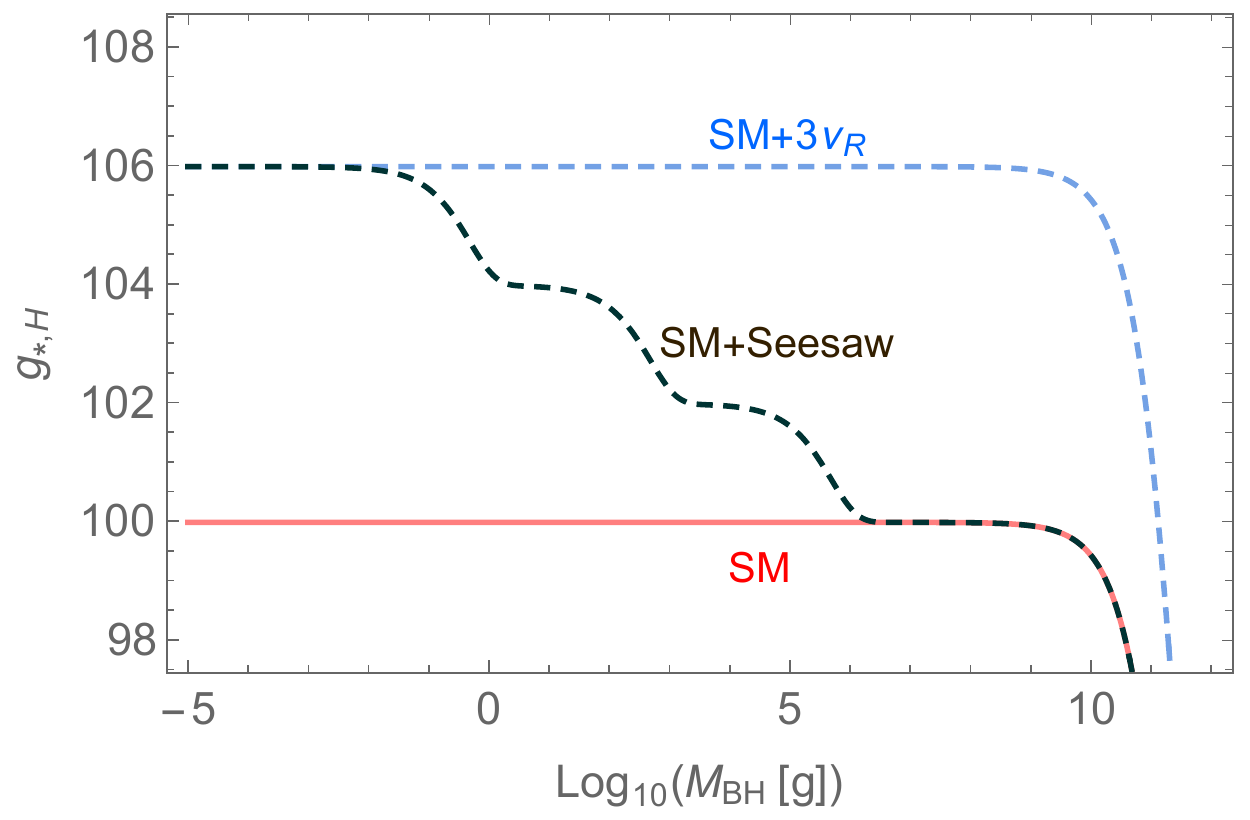}   \,\,\,\,
 \includegraphics[width=7.6cm]{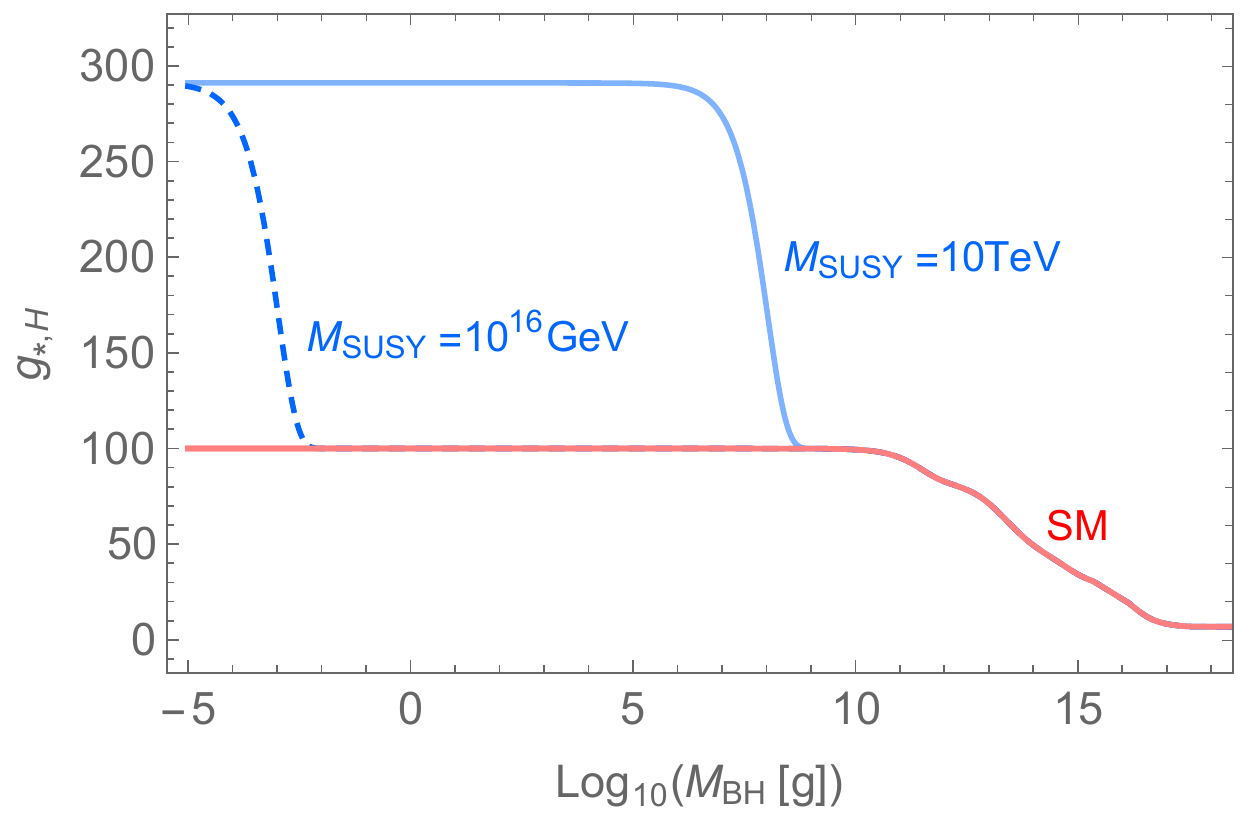}   
 \end{center}
\caption{\baselineskip=15 pt \small  
The function $g_{*,H}$ as a function of the BH mass for different models. Left: the SM compared to neutrino mass models, from bottom to top: the SM (including three left-handed neutrinos); the SM supplemented by a seesaw with hierarchical right-handed neutrino masses: $M_3=10^{14}$ GeV, $M_2=10^{11}$ GeV, $M_1=10^{8}$ GeV; 
the SM plus three light right-handed neutrinos (as in the Dirac mass case or in the case of a low energy seesaw).   
Right: the SM compared to supersymmetry realized at $10$ TeV and at $10^{16}$\,GeV. }
\label{fig-gsH}
\vskip .2 cm
\end{figure}

At BH temperatures well above the electroweak scale ($T_{BH} >>100$ GeV), corresponding to $M_{BH}<<10^{11}$ g, in the process of evaporation the full SM particle spectrum is emitted, the probability of each particle being given by its weight $g_{i,H}$. 
In this limit, for the full SM (with three left-handed neutrinos), $g_{*,H} = 100$. 
In general, the rate of mass loss by PBHs is enhanced if they emit not only SM particles but, if they exist, also heavy particles beyond the SM.
For instance, adding to the SM also three heavy right-handed neutrinos as in the seesaw mechanism, $g_{*,H} = 106$, as shown in the left panel of fig.\,\ref{fig-gsH}. In a supersymmetric model, the number of dofs at least triplicates\footnote{It might even become bigger including the dofs of the graviton and gravitino multiplets.}, as an effect of the lower spins of most of the supersymmetric particles; in the right panel of fig.\,\ref{fig-gsH} we show the impact on $g_{*,H}$, for the two cases of "low-energy" supersymmetry at $10$ TeV and of GUT-scale supersymmetry at $10^{16}$ GeV.

Ignoring the phenomenon of accretion, the lifetime of the BH is obtained by integrating eq.\,(\ref{eq-evf}) from the formation to the evanescence time, $t_{ev}$,
\beq
\tau= \int_{t_f}^{t_{ev}}    dt =   \frac{30720 \,\pi } {\mathcal{G} M_{Pl}^4}  \frac{\hbar c}{c^3}   \int^{M_{BH}}_0    
 \frac{M^2}{g_{\star, H} (T_{BH}) } \, d M\,
 = \frac{10640 \,\pi }{\mathcal{G} g_{\star, H} (T_{BH}) }  \frac{\hbar c}{c}   \frac{(M_{BH} c^2)^3}{(M_{Pl} c^2)^4}\,,
 \label{eq-tau}
\eeq
where in the last equality we assumed $g_{\star, H} (T_{BH})$ to be nearly constant during the BH lifetime (which is the case for the SM).
The integrand being proportional to $M^2$, the lifetime is determined by the first stages of evaporation, rather than by the last period. 

Since $\tau >> t_f$, we can approximate $t_{ev}\approx \tau$. The evanesce time is plotted as a function of the PBH mass  in fig.\,\ref{fig-TBHM}, assuming the SM dofs. One can see (as well known) that in order to preserve the good predictions of BBN, one needs $t_{ev} \lesssim 1$ sec, namely $M_{BH} \lesssim 10^9$ g.

The effect of accretion can be included by letting the mass loss be effective only after $t_{acc}$. 
Since $\tau >> t_{acc}$, the approximation $t_{ev} \approx \tau$ holds again.
The crucial difference is that in the integral in $dM$ of eq.\,(\ref{eq-tau}), the upper limit of integration should be replaced by $M^{acc}_{BH}$.  The BH lifetime including accretion, $\tau^{acc}$, becomes
\beq
\tau^{acc} \approx R_{acc}^3 \, \tau \,.
\eeq
For large $f_{acc}$, the enhancement of the lifetime can be quite large; the upper bound on $M_{BH}$ from BBN constraints would become accordingly stronger.

Assuming constant $g_{*,H}$, the variation of the BH mass with respect to time is simply  
\beq
M_{BH}(t) = M_{BH} \left( 1- \frac{t}{\tau} \right)^{1/3}\,.
\eeq
The first half of the BH mass is lost at $t=0.875 \tau$. The effect of accretion is easily incorporated in the formula above, giving
\beq
M^{acc}_{BH}(t) = R_{acc} M_{BH} \left( 1- \frac{t}{ R_{acc}^3 \tau} \right)^{1/3}\,.
\eeq

\subsection{The lifetime in BSM models}

We can split the SM and BSM contributions according to $g_{\star, H}  =g_{SM, H} +g_{BSM, H}$. The lifetime in BSM models gets enhanced according to
\beq
\tau^{BSM} = \frac{1}{  1 +\frac{g_{BSM, H} }{g_{SM,H}}}\, \tau^{SM} \approx  \left( 1 - \frac{g_{BSM, H}}{g_{SM,H}} \right) \,\tau^{SM}\,\,,
\eeq
where the last relation holds only for $g_{SM, H} >> g_{BSM, H} $.

We first discuss the effect of BSM particles lighter than about $10$ TeV. As can be seen from fig.\,\ref{fig-TBHM},
such particles would be produced in the evaporation of BHs with mass below about $10^9$ g.
Let consider in turn possible candidates.
For the massless graviton ($s=2$), $g_{BSM,H}= 0.10$, corresponding to a lifetime shortening of only $0.1\%$.
For $N_a$ axions ($s=0$), $g_{BSM,H}=1.82 \,N_a$; with $1$ and $10$ axions we the lifetime shortening would be by $2\%$ and $20 \%$ respectively.
Three additional very light right-handed neutrinos would provide $g_{BSM,H} =6$; a low energy seesaw or Dirac nature neutrinos would imply a lifetime shortening by $6 \%$.

In the case that supersymmetry is realized at about $10$ TeV (a scale potentially accessible to the LHC), we would have an enhancing of at least a factor of 3 in the dofs with respect to the SM; BHs lighter than $10^9$ g would then have a lifetime reduced by a factor of $1/3$ with respect to the SM.

Consider finally the case of particles heavier than $10$ TeV. This would be the case for: heavy right-handed seesaw neutrinos, GUT particles, supersymmetric GUT particles, monopoles, etc...  Such heavy particles would start to be emitted only when the BH temperature becomes larger than their mass. 
For instance, as can be seen from fig.\,\ref{fig-TBHM}, heavy particles with mass $10^{10}$ GeV are emitted if $M_{BH} \lesssim 10^3$ g. If the original BH mass was larger, these particles are emitted in the very last stages of evaporation, and their impact on the lifetime is marginal. Otherwise their impact on the lifetime might be significant: as an example, three heavy right-handed neutrinos with mass $10^{10}$ GeV imply a shortening of the BH lifetime by factor $6\%$ for $M_{BH} \lesssim 10^3$ g. 
GUT-scale supersymmetric particles would be emitted only when $M_{BH} \lesssim 10^{-3}$ g; since the initial BH mass has to be larger than $0.1$ g, there is no significant effect on the BH lifetime.

\section{BH abundance dynamics}

According to the first Friedmann equation, a universe that contains both radiation and BHs satisfies
\beq
H^2 = \frac{8 \pi G}{3} \left(  \rho_{R} +  \rho_{BH}   \right)\,,
\eeq
where $\rho_R$ and $\rho_{BH}$ are the radiation and BH mass densities at time $t$, which evolve as:
\beq
\frac{d\rho_R}{dt} +4 H \rho_R =-\frac{dM_{BH}/dt}{M_{BH}}\rho_{BH}
\eeq
\beq
\frac{d\rho_{BH}}{dt} +3 H \rho_{BH} =   \frac{dM_{BH}/dt}{M_{BH}}\rho_{BH}\,.
\eeq
Since $\rho_{BH} \propto a^{-3}$, while $\rho_R \propto a^{-4}$, one obtains
\beq
f(t) =\frac{\rho_{BH}(t)}{\rho_{R}(t)} \propto a(t)\,.
\eeq

\begin{figure}[t!]
\vskip .0 cm 
\begin{center}
 \includegraphics[width=8.3cm]{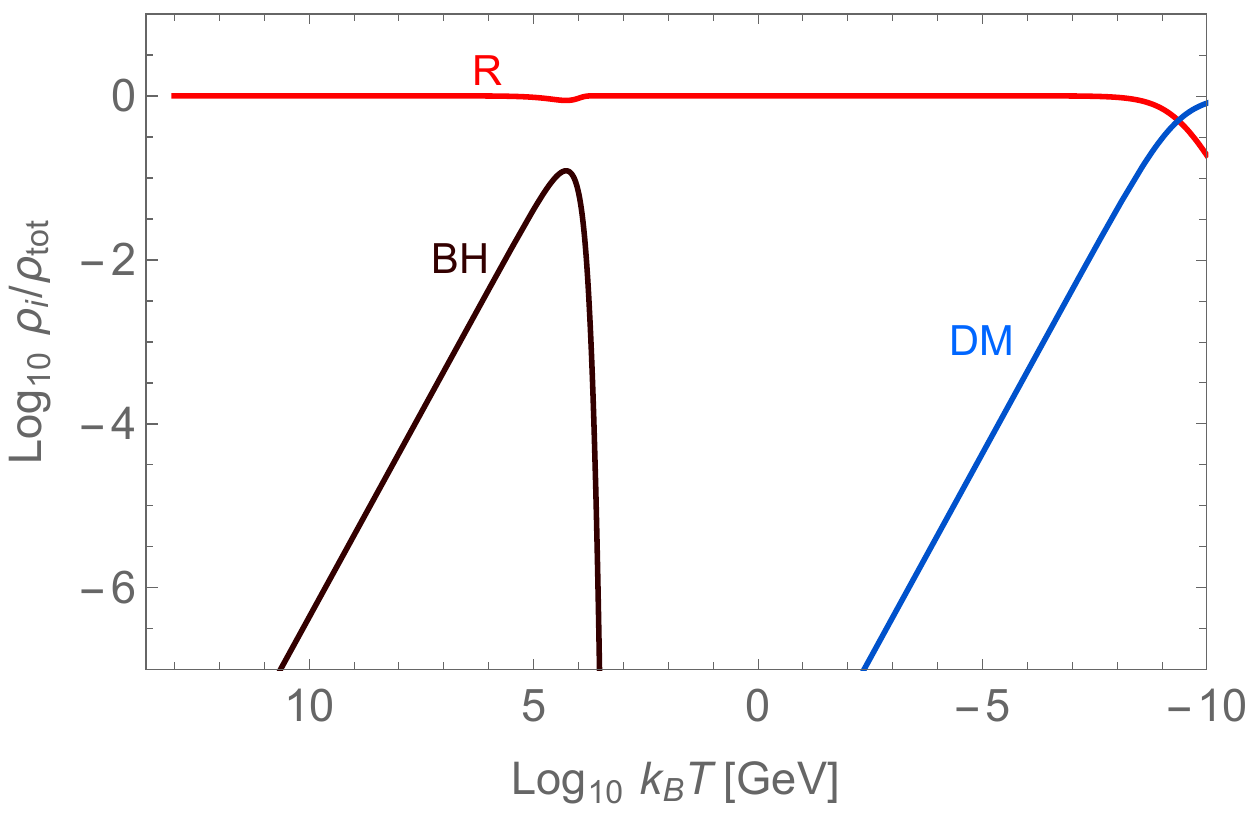}   \\ \vskip 1 cm
   \includegraphics[width=8.3cm]{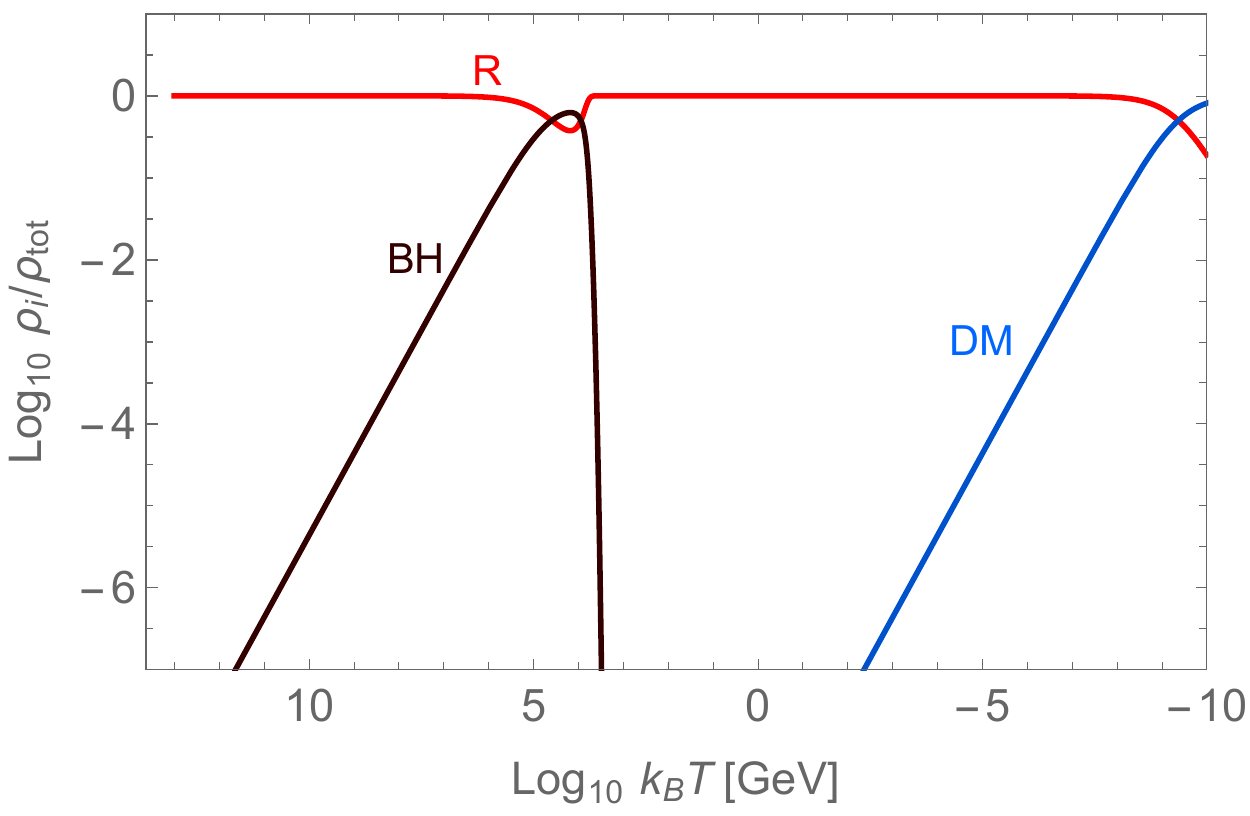}    \\ \vskip 1 cm
  \includegraphics[width=8.3cm]{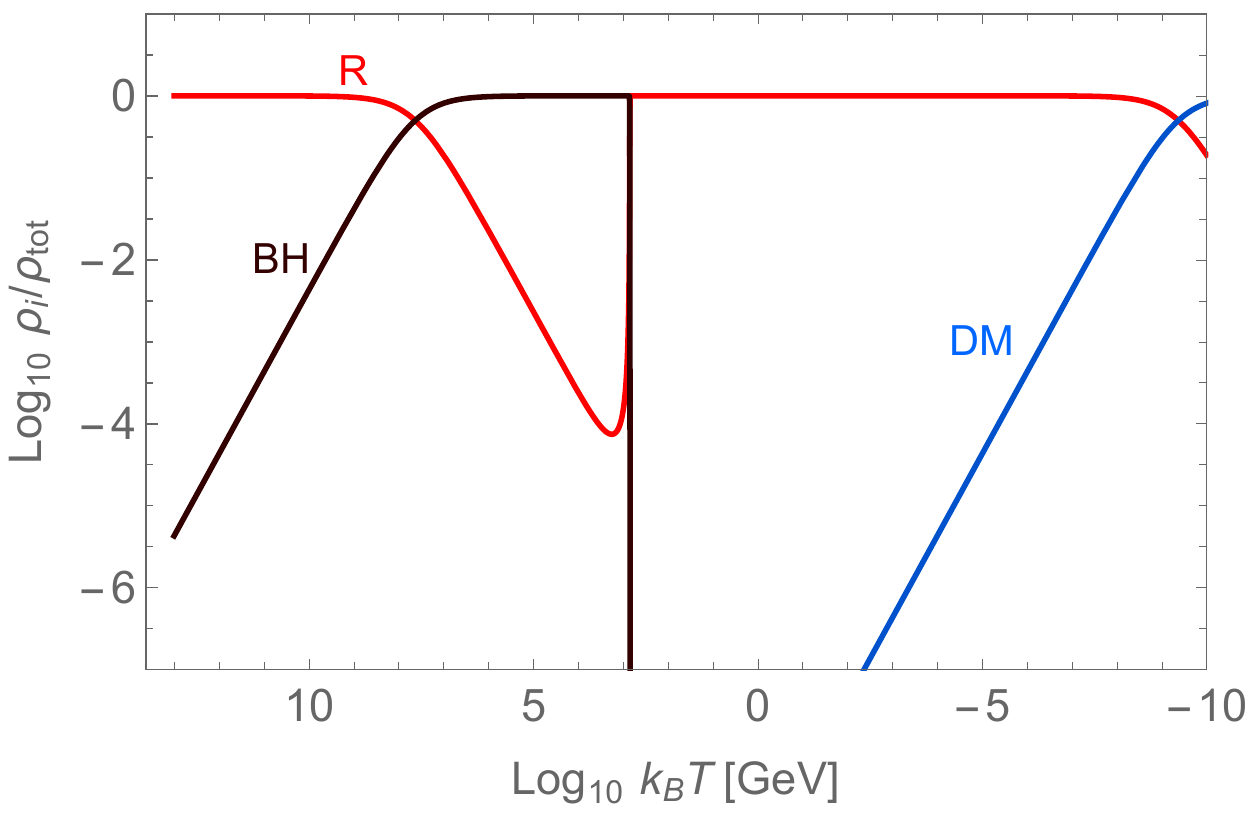} 
 \end{center}
\caption{\baselineskip=15 pt \small  Taking $M_{BH}=10^4$ g, so that $k_B T_f \sim 10^{13}$ GeV while 
$k_B T_{ev} \sim 10^{4}$ GeV (see fig.\,\ref{fig-TBHM}), we show the fractions of the energy densities in radiation (R), PBHs (BH) and dark matter (DM), as a function of the BH mass.
Upper: radiation domination, $\beta = 10^{-10}$. Middle: evanescence occurs just at BH domination, $\beta =10^{-9}$. Lower: evanescence occurs after a long period of BH domination, $\beta  = 10^{-6}$. }   
\label{fig-rho1}
\vskip .0 cm
\end{figure}

Ignoring the effect of accretion for simplicity, we start the evolution at the formation time $t_f$, when the universe has the temperature $T_f$.
Depending on the parameter $\beta=  f(t_f)$, it might happen that PBHs come to dominate the energy content of the universe at some time $t_{BH}$ (PBHs should anyway evaporate before BBN in order not to destroy its good predictions);
this scenario is called BH domination for short. The scenario in which BHs never dominate the energy content of the universe is rather referred to as radiation domination.

Due to the evaporation of PBHs, entropy is not conserved
\beq
\frac{ds}{dt} +  3 H s = -\frac{dM_{BH}/dt}{M_{BH}}  \frac{\rho_{BH}}{T}\,,
\eeq
which in terms of the relation between temperature and times reads (we assume that the entropic dofs are constant)
\beq
\frac{dT}{dt} \frac{1}{T} = - \left(  H  + \frac{dM_{BH}/dt}{M_{BH}}  \frac{\rho_{BH}}{4 \rho_R} \right) = - H_{eff}\,.
\label{eq-Tt}
\eeq
Only in the last stages of evaporation the value of $H_{eff}$ might differ significantly from $H$.
It is useful to exploit this latter equation to rewrite the system above
\beq
\frac{d\rho_R}{dT}  + \frac{4}{T}  \rho_R =0
\eeq
\beq
\frac{d\rho_{BH}}{dT}  + \frac{3}{T} \frac{H}{H_{eff}} \rho_{BH} = -\frac{dM_{BH}/dt}{M_{BH}} \frac{1}{H_{eff} T}  \rho_{BH} \,.
\eeq

As an illustrative example, in fig.\,\ref{fig-rho1} we show the evolution of $\rho_R$ and $\rho_{BH}$ as a function of the temperature, assuming $M_{BH}=10^4$ g and, from top to bottom, $\beta=10^{-10}, 10^{-8}, 10^{-6}$. 
The associated formation temperature is $k_B T_f \sim 10^{13}$ GeV, see fig.\,\ref{fig-TBHM}.
In the top panel the universe is always radiation dominated; in the middle panel there is a short epoch when $\rho_R \sim \rho_{BH}$; in the bottom panel there is a significantly long period of BH domination, ending when BHs eventually evaporate, which happens at $k_B T_{ev}\sim 10^4$ GeV (before BBN).
We also show the density of dark matter, $\rho_{DM}$, which has to equal radiation at matter-radiation equality, $t_{EQ}$.

We now study in some detail the temperature of the radiation bath at the evanescence of PBHs.

\subsection{Temperature at evanescence}

Consider first the case that the universe is still radiation dominated at $t_{ev}$,
then 
\beq
\frac{8 \pi G}{3} \rho_R(t_{ev})
=H^2(t_{ev}) \approx \frac{1}{4 \tau^2} \,\,\, .
\eeq  
Using eq.\,(\ref{eq-rhoRT}) in the expression above, we get the temperature of the radiation for radiation domination
\beq
k_B T_{ev} 
=   \left( \frac{ 45 } { 16 \pi^3  g_*(T_{ev})  } \right)^{1/4}  
 \left( \frac{ \hbar \, (M_{Pl} c^2) }{  \tau} \right)^{1/2}
 =   \left( \frac{ 45 } { 16 \pi^3  g_*(T_{ev})  } \right)^{1/4}  
\left(  \frac{\mathcal{G} g_{\star, H} (T_{BH}) } {10640 \,\pi }  \right)^{1/2}
 \frac{(M_{Pl} c^2)^{5/2}}{(M_{BH} c^2)^{3/2}} \,\,.
\eeq
The value of $k_B T_{ev}$ is shown in fig.\,\ref{fig-TBHM}. The effect of accretion is to lower the evanescence temperature according to: $T^{acc}_{ev}= R_{acc}^{-3/2} T_{ev}$.

It is useful to write explicitly the ratio
\beq
\frac{T_{f}}{T_{ev}} = \gamma^{1/2} \left(  \frac{g_*(T_{ev})}{g_*(T_{f})}\right)^{1/4} \left( \frac{10640 \,\pi}{{\mathcal G} \, g_{*,H}(T_{BH})} \right)^{1/2} \frac{M_{BH}}{M_{Pl}}\,.
\label{eq-TevTf}
\eeq

In the case of a sufficiently long period of BH dominance before evanescence, namely $ t_{BH} << t_{ev} $,
we instead have
\beq
\frac{8 \pi G}{3} \rho_{BH}(t_{ev})
=H^2(t_{ev}) \approx \frac{4}{9 \tau^2} \,\,\, .
\label{eq-FBHd}
\eeq  
Since we can grossly estimate that the energy density in BHs goes into radiation after evanescence, $\rho_{R}(t_{ev}^+) \approx \rho_{BH}(t_{ev}^-)$, the temperature in the BH dominated case is slightly higher than in the radiation dominated one
\beq
 T^{BH}_{ev}  =\frac{2}{ \sqrt{3} }  T_{ev}   \approx 1.15 \,  T_{ev} 
 \,\,.
\eeq
The increase is then about $15\%$ with respect to the radiation dominated case. 

As an effect of the modification to $H$ in eq. (\ref{eq-Tt}), the radiation temperature undergoes also a small reheating in the last stages of the BH lifetime, estimated to be about $7$ GeV for $M_{BH}=2\times 10^3$ g in ref.\,\cite{Majumdar:1995yr}.
In the following we neglect this effect.

\subsection{Radiation vs BH domination at evanescence}

\begin{figure}[t!]
\vskip .2 cm 
\begin{center}
 \includegraphics[width=10cm]{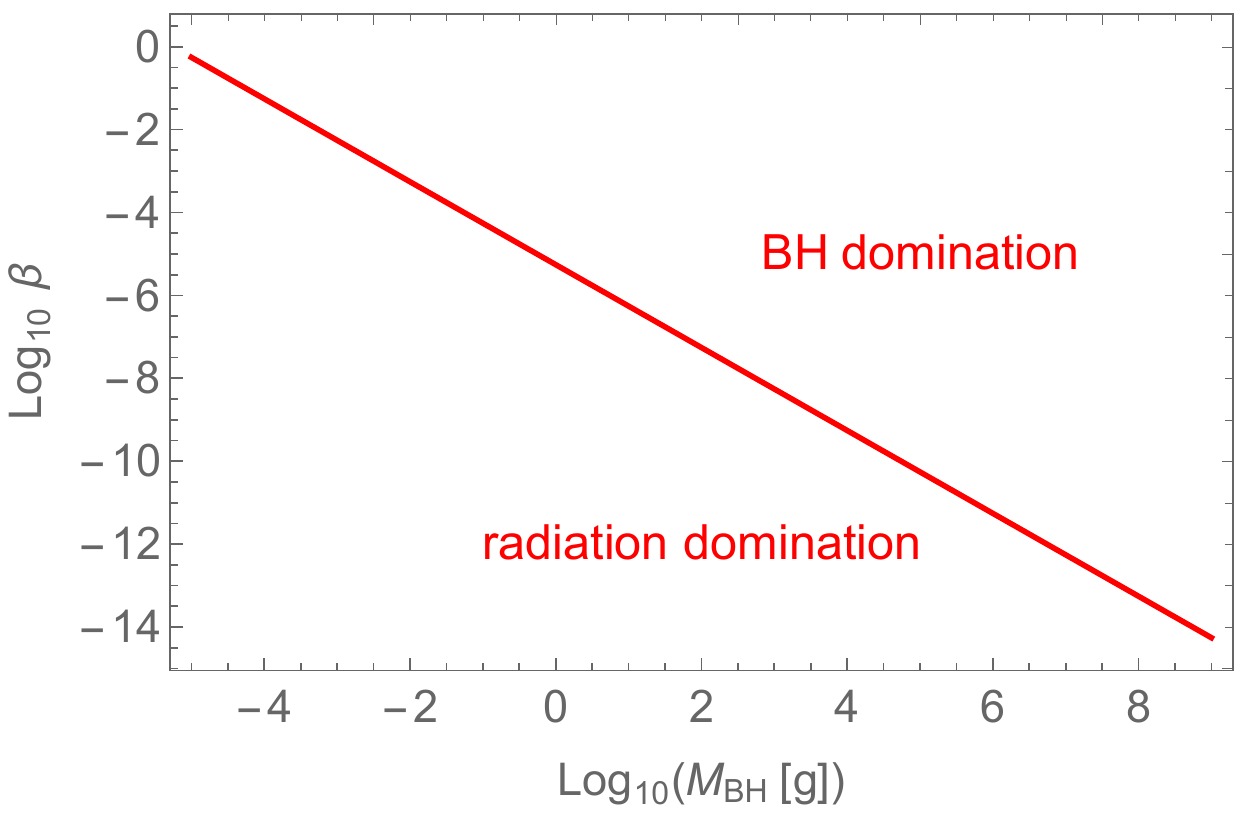}   
 \end{center}
\caption{\baselineskip=15 pt \small  
The values of $\beta$ providing radiation or BH dominance, as a function of the BH mass.}   
\label{fig-BHdom}
\vskip .2 cm
\end{figure}

If PBHs completely evaporate when the universe is still radiation dominated, 
\beq
f(t_{ev}) = \frac{\rho_{BH}(t_{ev})}{\rho_R(t_{ev})}  <1\, .
\eeq
In the case of radiation domination, neglecting the time variation of $M_{BH}$, 
\beq
\frac{f(t_{ev})}{f(t_f)} =  \frac{\rho_{BH}(t_{ev}) }  {\rho_{BH}(t_{f}) }    \frac {\rho_{R}(t_f)} {\rho_{R}(t_{ev})} 
=  \frac {a(t_{ev})} {a(t_{f})}  =  \left( \frac {t_{ev}}{t_{f}}   \right)^{1/2} 
=  \left( \frac{g_*(T_f)}{g_*(T_{ev})} \right)^{1/4} \frac {T_{f}}  {T_{ev}} \,.
\label{eq-ftevftf}
\eeq
Using the two equations above and eq.\,(\ref{eq-TevTf}), the condition of evanescence during radiation dominance becomes 
\beq
\beta= {f(t_f)}    < \bar \beta =\left( \frac{g_*(T_{ev})}{g_*(T_{f})} \right)^{1/4}  \frac  {T_{ev}} {T_{f}} 
=  \gamma^{-1/2}  \left( \frac{\mathcal{G} g_{\star, H} (T_{BH}) } {10640 \,\pi } \right)^{1/2} \frac{M_{Pl} }{M_{BH}}   \,\,,
\label{eq-beta}
\eeq
which is shown in fig. \ref{fig-BHdom}. Numerically, the factor $( g_*(T_{ev}) / g_*(T_{f}))^{1/4}$ ranges from $0.5$ to $1$ in the SM, and is close to 1 for $k_B T_{ev}\gtrsim 170$ GeV, namely $M_{BH}\lesssim 5.5 \times 10^5$ g.
If this condition is not satisfied, BHs evaporate during BH domination.
The three cases considered in fig.\,\ref{fig-rho1} are consistently reproduced in the region plot of fig. \ref{fig-BHdom}.

Note also that $\bar \beta$ gets lower including accretion, $\bar \beta^{acc}= R_{acc}^{-3/2} \bar \beta$.

\subsection{BH abundance at evaporation for radiation domination}

Let $Y_{BH}(t)$ be the number-to-entropy density of BHs at time $t$ 
\beq
Y_{BH}(t) =  \frac{n_{BH}(t)}{s(t)}
=\frac{1}{M_{BH}(t) }\frac{\rho_{BH}(t)}{s(t)} 
= \frac{1}{M_{BH}(t) } f(t) \frac{\rho_{R}(t)}{s(t)} \, ,
\label{eq-YBH}
\eeq
where the entropy density is defined as 
\beq
s(t) = \frac{ 2 \pi^2 g_{*,S}(T) }{45} \frac{(k_B T)^3}{(\hbar c )^3} \,.
\label{eq-s}
\eeq
The difference between $g_{*}(T)$ and $g_{*,S}(T)$ can in general be neglected. 

Assuming radiation domination and neglecting the time dependence of $M_{BH}$, 
we can calculate $Y_{BH}(t_{ev})$ using eqs.\,(\ref{eq-YBH}), (\ref{eq-s}) and (\ref{eq-rhoRT}),
\beq
Y_{BH}(t_{ev})
=   f(t_{ev})  \frac{ 3 }{4}   \frac{g_*(T_{ev})}{g_{*,S}(T_{ev})} \frac{k_B T_{ev}}{M_{BH} c^2}  
\,,
\label{eq-Ytev}
\eeq
and, in the same way, 
\beq
Y_{BH}(t_f)
=  f(t_f)   \frac{ 3 }{4} \frac{g_*(T_{f})}{g_{*,S}(T_{f})}    \frac{k_B T_{f}}{M_{BH} c^2} 
\,.
\eeq
Using eq.\,(\ref{eq-ftevftf}), the relation between $Y_{BH}(t_{ev})$ and $Y_{BH}(t_{f})$ is then 
\beq
\frac{1}{\alpha_{BH}}=\frac{Y_{BH}(t_{ev}) }{Y_{BH}(t_{f})}
=  \left( \frac{g_*(T_{ev})}{g_*(T_{f})} \right)^{3/4} \frac{g_{*,S}(T_{f})}{g_{*,S}(T_{ev})}  
\approx \left( \frac{g_*(T_{f})}{g_*(T_{ev})} \right)^{1/4}  \,.
\label{eq-alfaBH}
\eeq
where $\alpha_{BH}$ encodes the non conservation of entropy during the BH lifetime: $\alpha
_{BH} (a^3 s)_{f}=(a^3 s)_{ev}$.

Explicitly, using eqs.\, (\ref{eq-Ytev}), (\ref{eq-Tf}) and assuming $g_*(T_{f}) \approx g_{*,S}(T_{f}) $,
\beq
Y_{BH}(t_{ev}) 
\approx \frac{1}{\alpha_{BH}} 
\beta \frac{ 3 }{4} 
 \left(  \frac{45 \gamma^2  } {16 \pi^3 g_*(T_{f}) }   \right)^{1/4}  \left( \frac{M_{Pl}  }{M_{BH}} \right)^{3/2} \,.
 \label{eq-YBHRev}
\eeq

Including the effect of accretion, $Y_{BH}^{acc}(t_{ev}) = R_{acc}^{-5/2} Y_{BH}(t_{ev})$.

\subsection{BH abundance at evaporation for BH domination}

For BH domination, there is no $\beta$ dependence. 
We follow the argument of ref. \cite{Baumann:2007yr} in order to calculate 
$n_{BH}(t_{ev}) \approx \rho_{BH}(t^-_{ev})/M_{BH}$.  
 For $t_{ev} >> t_{BH}$, using the first Friedmann equation for BH domination, eq.\,(\ref{eq-FBHd}), one has
\beq
n_{BH}(t_{ev})  
=  \frac{1}{6 \pi }  \frac{(M_{Pl} c^2)^2}{ (\hbar c) \,(c\,\tau )^2  (M_{BH} c^2)}  
=  \frac{1}{6 \pi } 
\left( \frac{\mathcal{G} g_{\star, H} (T_{BH}) }{10640 \,\pi } \right)^2    
\frac{(M_{Pl} c^2)^{10}}{(\hbar c)^3 \,(M_{BH} c^2)^7}\,.
\eeq
Since the entropy at evaporation is
\beq
s(t_{ev})= \frac{ 2 \pi^2 g_{*,S}(T_{ev}) }{45} \frac{(k_B T_{ev})^3}{(\hbar c)^3}
=\frac{ 2 \pi^2 g_{*,S}(T_{ev}) }{45} \left(   \frac{ 5   } {  \pi^3  g_*(T_{ev})   } \right)^{3/4}
\left(  \frac{ M_{Pl} c^2}{(\hbar c) \,(c\, \tau)} \right)^{3/2} \,,
\eeq
we obtain the explicit expression for BH domination:
\beq
Y_{BH}(t_{ev}) =  \frac{n_{BH}(t_{ev})}{s(t_{ev})} 
 = \frac{15} { 4 \pi^3 g_{*,S}(T_{ev}) }  \left(   \frac{  \pi^3  g_*(T_{ev})   }{ 5   }  \right)^{3/4}
 \left( \frac{\mathcal{G} g_{\star, H} (T_{BH}) } {10640 \,\pi } \right)^{1/2}
 \left( \frac{M_{Pl} }{M_{BH}} \right)^{5/2} \,.
 \label{eq-YBHBHev}
\eeq

Also in this case, including the effect of accretion, $Y_{BH}^{acc}(t_{ev}) = R_{acc}^{-5/2} Y_{BH}(t_{ev})$.

\section{Particle production by PBHs}

Let $N_{X}$ be the number of X particles produced in the evaporation of a single BH. 
We first calculate $N_X$ following Baumann et al. \cite{Baumann:2007yr}, but generalizing to the case of Planck scale relics. 
Secondly, assuming that the $X$ particle is stable, we determine whether it might contribute to DR. We also revisit the lower limits on the mass of a stable X particle emitted by a BH, in the case it constitutes the dominant part of the DM.

\subsection{The number of X particles produced by a single BH}

The definition of the Hawking temperature implies the differential mass decrease
\beq
-dE=d(M_{BH} c^2)= -\frac{1}{8 \pi}  \frac{(M_{Pl} c^2)^2}{(k_B T_{BH})^2} d(k_B T_{BH})\,,
\eeq
where $dE$ is the energy emitted by the BH while loosing the mass $dM_{BH}$.
Since the radiated particles have mean energy $3 k_B T_{BH}$, the differential number of particles emitted is
\beq
dN=\frac{dE}{3 k_B T_{BH}} 
= \frac{(M_{Pl} c^2)^2}{24 \pi}  \frac{d(k_B T_{BH})}{ (k_B T_{BH})^3} 
\,.
\eeq
The number of the X particles is obtained by properly accounting for the dofs and by integrating the above expression
\beq
N_X
= \frac{g_{X,H}}{g_{*,H}} \int_0^{N} dN \,,
\label{eq-NX}
\eeq
where $g_{X,H}$ are the X particle dofs, while $g_{*,H}$ are the dofs of all emitted particles.

Consider first the case of "light" X particles, that is $M_X c^2 < k_B T_{BH}$. BHs start emitting the X particles immediately, when their mass is $M_{BH}$, and stop when they reach the relic mass $M_{BH}^R << M_{BH}$. 
Defining the relic BH temperature as $ (k_B T_{BH}^R ) = 1/(8 \pi)  (M_{Pl} c^2)^2 /(M^R_{BH} c^2)$, 
we integrate eq.\,(\ref{eq-NX}) obtaining
\beq
N_X 
= \frac{g_{X,H}}{g_{*,H}} \,   \frac{(M_{Pl} c^2)^2}{24 \pi}      \int_{k_B T_{BH}}^{k_B T_{BH}^R}  \frac{d(k_B T)}{ (k_B T)^3}
= \frac{g_{X,H}}{g_{*,H}} \frac{4 \pi}{3} 
\left(  \frac{M_{BH}}{ M_{Pl}}\right)^2 \left(1- \left( \frac{M_{BH}^R}{M_{BH}} \right)^2 \right) \,.
\eeq
Note that the most important contribution to the integral comes from the smaller values of the temperature, corresponding to 
the higher values of the BH mass.
Assuming BHs completely evaporates (no relics),
\beq
N_X 
= \frac{g_{X,H}}{g_{*,H}} \frac{4 \pi}{3} \left(  \frac{M_{BH}}{ M_{Pl}}\right)^2\,\,.
\label{eq-NXl}
\eeq 
In the case of "light" X particles, the number of emitted particles depends just on $M_{BH}$. 

In the case of "heavy" X particles,  $M_X c^2 > k_B T_{BH}$. Assuming no relics, the production proceeds from the moment when the BH  temperature $k_B T_{BH}$ goes below $M_X c^2$, thus:
\beq
N_X = \frac{g_{X,H}}{g_{*,H}} \, \frac{(M_{Pl} c^2)^2}{24 \pi}   \int_{M_{X} c^2}^{k_B T_{BH}^R}  \frac{d(k_B T)}{ (k_B T)^3} 
= \frac{g_{X,H}}{g_{*,H}}  \left(  \frac{1}{48 \pi} 
\left(  \frac{M_{Pl}}{ M_{X}}\right)^{2} - \frac{4 \pi}{3}  \left( \frac{M_{BH}^R}{M_{Pl}} \right)^2  \right)\,\,.
\eeq 
Assuming BHs completely evaporates (no relics),
\beq
N_X = \frac{g_{X,H}}{g_{*,H}}   \frac{1}{48 \pi} \left(  \frac{M_{Pl}}{ M_{X}}\right)^{2}\,\,.
\label{eq-NXh}
\eeq 
In the case of "heavy" X particles, which are emitted only in the final stages of the BH lifetimes, the number of emitted particles depends only on $M_X$. This is the case for GUT scale ($10^{15}$ GeV) particles, see figs.\,\ref{fig-TBHM} and \ref{fig-gsH}: they start to be produced when the mass of the BH goes below about $10^{-2}$\,g.

\begin{figure}[t!]
\vskip .2 cm 
\begin{center}
 \includegraphics[width=10cm]{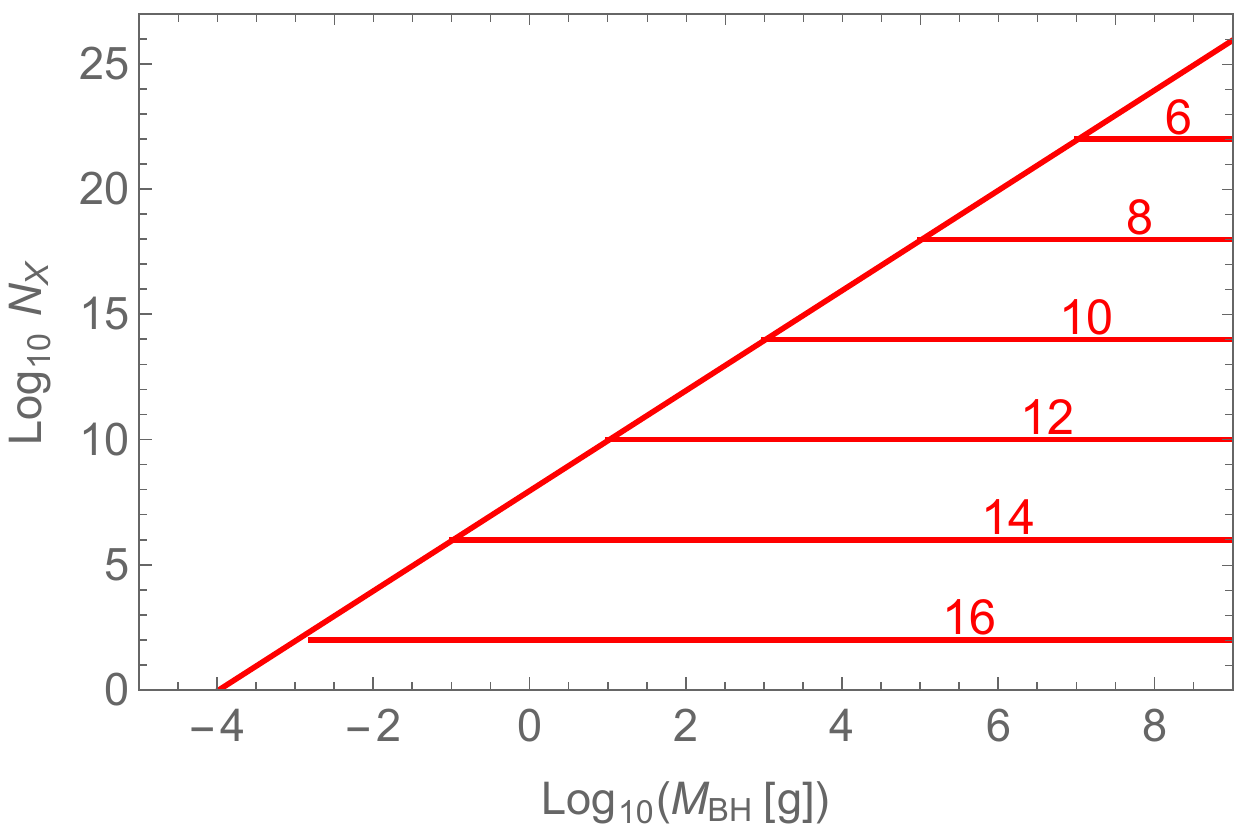}   
 \end{center}
\caption{\baselineskip=15 pt \small  
$N_X$ as a function of the BH mass assuming $g_{X,H}/g_{*,H}=1/100$, and for various values of ${\rm Log}_{10} (M_X c^2[ {\rm GeV}])$, as indicated.}   
\label{fig-NX}
\vskip .2 cm
\end{figure}

In fig.\,\ref{fig-NX} we show $N_X$ as a function of the BH mass, taking $g_{X,H}/g_{*,H}=1/100$, and for various values of ${\rm Log}_{10} (M_X c^2[ {\rm GeV}])$, as indicated. For instance, for $M_X = 10^{10}$ GeV, we can see that the "light" regime applies for the range $M_{BH}=0.1-10^3$ g, while the "heavy" regime for  $M_{BH}=10^3-10^9$ g.

The effect of accretion is included by replacing $M_{BH}$ with $M^{acc}_{BH}$ in the above expressions.
As a consequence, in the "light" case, $N_X^{acc}= R_{acc}^2 N_X$, while in the "heavy" case there is no change.

\subsection{Stable particles as dark matter and dark radiation}

If the particles X produced in the evaporation of PBHs are stable, they might significantly contribute to DM and, if sufficiently light, also to DR \cite{Lennon:2017tqq,Hooper:2019gtx}.
Here we review the argument of ref.\,\cite{Hooper:2019gtx}, generalizing it to include the effect of entropy non conservation.

In order for X particles to contribute towards DR, their average kinetic energy evaluated at
$t_{EQ}$ must exceed their mass: $p_{EQ} c\approx \langle E(t_{EQ}) \rangle \gtrsim M_X c^2$. 
The average kinetic energy of the emitted particles is approximately\,\cite{Fujita:2014hha}
\beq
p_{ev} c \approx \langle E (t_{ev}) \rangle \approx \frac{1}{N} \int_0^{N} 3 k_B T_{BH} \,dN
= 6 (k_B T_{BH})^2   \int_{k_BT_{BH}}^{\infty}    \frac{d(k_B T)}{ (k_B T)^2} 
= 6 \, (k_B T_{BH})\,.
\eeq
The more refined calculation of ref.\,\cite{Lennon:2017tqq} (that we verified finding full agreement) gives a numerical factor smaller than the factor 6 in the right hand side of the equation above: this factor was calculated to be about $1.3$ (or even smaller, if the effect of the expansion of the universe during the BH lifetime is taken into account). To be conservative, we define the parameter $\delta$, such that $\langle E (t_{ev}) \rangle  \approx \delta k_B T_{BH}$, and in the following discussion we adopt as reference value $\delta=1.3$.
Since the momentum scales as the scale factor, 
\beq
\langle E (t_{EQ}) \rangle 
\approx \langle E (t_{ev}) \rangle \, \frac{a_{ev}}{a_{EQ}}
=  \delta \, (k_B T_{BH})\frac{1}{\alpha'} \frac{T_{EQ}}{T_{ev}} \left( \frac{g_{*,S}(T_{EQ})}{g_{*,S}(T_{ev})} \right)^{1/3} \,\, , 
\label{eq-ETEQ}
\eeq
where in the last equality we assumed entropy conservation from evaporation to matter-radiation equality, $\alpha' (s a^3)_{ev}=(s a^3)_{EQ}$. 

\begin{figure}[h!]
\vskip .2 cm 
\begin{center}
 \includegraphics[width= 10.2cm]{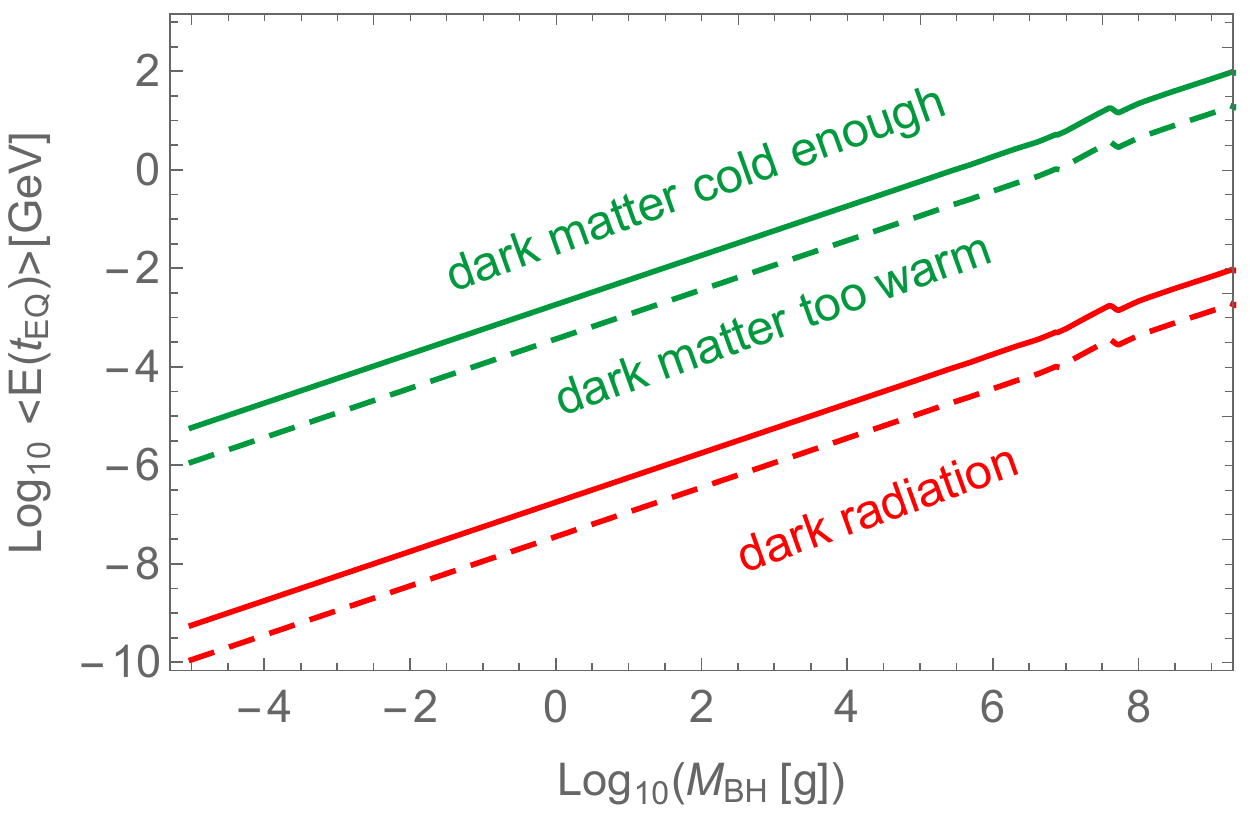}  \,\,\, 
 \end{center}
\caption{\baselineskip=15 pt \small  
(Red) lower solid ($\alpha'=1$) and dashed ($\alpha'=5$) curves: $\langle E (t_{EQ}) \rangle$ as a function of the BH mass; stable particles with mass below $\langle E (t_{EQ}) \rangle$ contributes to DR.
(Green) upper solid ($\alpha'=1$) and dashed ($\alpha'=5$) curves: the value of $m_X c^2$ below which the X particles are too warm to be the main component of the DM. The plot is obtained taking for definiteness $k_B T_{EQ} = 0.75$ eV and $\delta=1.3$.}   
\label{fig-EDR}
\vskip .2 cm
\end{figure}

We show $\langle E (t_{EQ}) \rangle$ in fig.\,\ref{fig-EDR}, taking $k_B T_{EQ} \approx 0.75$ eV, $\delta=1.3$, $\alpha'=1$ (solid) and  $\alpha'=5$ (dashed). With $\alpha'=1$, in order to contribute to DR, X particles must be lighter than about $0.2$ keV, $20$ keV, $2$ MeV, with BHs weighting respectively $1,10^4,10^8$ g. For $\alpha'>1$, the upper bound gets stronger by a factor $1/\alpha'$.

The effect of accretion is easily incorporated by considering that $T_{BH} /T_{ev} \propto   M^{1/2}_{BH}$, so that 
\beq
\langle E^{acc} (t_{EQ}) \rangle  = R_{acc}^{1/2} \,\langle E (t_{EQ}) \rangle \,.
\eeq
As opposite to the non conservation of entropy, it goes in the direction of enhancing $\langle E (t_{EQ}) \rangle$.

Particles contributing to DR include for instance: for $s=0$, axions, majorons; for $s=1/2$, neutralinos, axinos, sufficiently long-lived light right-handed (sterile) neutrinos (not completely stables as they might decay into three left-handed neutrinos or into a photon and a left-handed neutrino); for $s=3/2$, gravitinos; for $s=2$, gravitons.

\subsection{Constraints on warm DM}
 \label{sec-wdm}

If the X particles is going to provide the full contribution to DM, one has to check that it was cold enough not to waste structure formation. 
The X particles are emitted with a distribution of momenta. Nevertheless, a simple argument based on mean quantities, 
as discussed in Fujita et al. \cite{Fujita:2014hha} (see also \cite{Lennon:2017tqq} for a more refined approach to this issue),
allows to derive an order of magnitude estimate for the lower value of $m_X$ that would be compatible with structure formation.

The momentum of the X particle is red-shifted by the expansion of the universe,
\beq
 p_{now} = \frac{a_{ev}}{a_{now}} p_{ev} = a_{EQ}  \frac{a_{ev}}{a_{EQ}} \frac{\langle E (t_{ev}) \rangle}{c}
 =  \frac{\Omega_R}{\Omega_{M}}   \frac{\langle E (t_{EQ}) \rangle}{c} 
\eeq
where we used eq.\,(\ref{eq-ETEQ}), $a_{now}=1$, and $a_{EQ}=\Omega_R/\Omega_{M} \approx 1.8\times 10^{-4}$. 
Assuming that it is no more relativistic, the velocity of the X particle now is then
\beq
 \frac{v_X}{c}= \frac{p_{now}}{c \,M_X} =  1.8\times 10^{-4} \frac{\langle E (t_{EQ}) \rangle}{M_X c^2} \,.
 \label{eq-vX}
\eeq
In the case of significant accretion, one would obtain larger velocities, 
$\frac{v_X}{c}|_{acc}= R_{acc}^{1/2}\frac{v_X}{c}$.

The lower bound on the mass of a thermal early decoupled warm DM candidate, can be translated into a lower bound on the present velocity of a generic warm DM candidate \cite{Fujita:2014hha}. 
Here we update the argument of ref. \cite{Fujita:2014hha} (based on \cite{Viel:2005qj}), with the new data from ref. \cite{Irsic:2017ixq}.
Assume that the warm DM was relativistic at decoupling. Since the velocity scales as the scale factor, and assuming entropy conservation from decoupling to the present epoch,
\beq
 \frac{v_{WDM} }{c} =a_{dec} \approx  \left( \frac{g_{*,S}(T_{now})}{g_{*,S}(T_{dec})}\right)^{1/3} \frac{k_B T_{now}}{m_{WDM} c^2} 
 \lesssim 1.8 \times 10^{-8} \,,
\label{eq-limV}
\eeq
where in the last relation we took $k_B T_{dec} \approx m_{WDM} c^2 \gtrsim 3.5$ keV (at 2$\sigma$) \cite{Irsic:2017ixq}.

If X particles are going to fully contribute to DM, we have to require $v_X\lesssim v_{WDM}$. Combining eqs.\,(\ref{eq-limV}) and (\ref{eq-vX}), one obtains a lower bound on the mass of the X particle 
\beq
M_X c^2   \gtrsim 
 10^4  \,\langle E (t_{EQ}) \rangle \,.
\eeq
As shown in fig.\,\ref{fig-EDR}, with $\alpha'=1$, the X particles must be heavier than about $2 \times 10^{-3},0.2, 20$ GeV, for PBHs weighting respectively $1,10^4,10^8$ g. If $\alpha' >1$, the lower bound on the X mass get relaxed by a factor $1/\alpha'$.
On the contrary, if accretion plays a significant role, the lower bound on $M_X$ gets stronger by a factor $R_{acc}^{1/2}$.

\section{Stable particles as dark matter}

One can treat evaporation \cite{Baumann:2007yr} as if all particles were produced at a single instant, $t\approx t_{ev}$. 
The present number-to-entropy density of a stable particle $X$ produced by evaporation is directly related to the BH abundance at evaporation\footnote{Instead, in ref. \cite{ Morrison:2018xla}, dealing with the radiation dominated case, the formation time is used: this is not justified as there might be entropy variation from the formation to the evaporation time, see eq. (\ref{eq-alfaBH}).}
 \cite{Baumann:2007yr, Fujita:2014hha} 
\beq
Y_X(t_{now}) =\frac{n_{X}(t_{now}) }{s(t_{now}) }= \frac{1}{\alpha}\frac{n_{X}(t_{ev}) }{s(t_{ev}) }
= \frac{1}{\alpha} N_{X} \frac{n_{BH}(t_{ev})}{s(t_{ev})} = \frac{1}{\alpha} N_{X}  Y_{BH}(t_{ev})\,,
\label{eq-YXtnow}
\eeq
where $\alpha$ parametrizes a possible entropy production after evanescence, $\alpha (s a^3)_{ev}=(s a^3)_{now}$, 
$N_{X}$ is the number of X particles produced in the evaporation of a single BH. It is reasonable to assume that entropy is conserved from matter-radiation equality to the present time, that is $\alpha \approx \alpha'$.

The cosmological abundance now is 
\beq
\Omega_{X} =\frac{\rho_{X}   }{\rho_{c}}=\frac{M_{X}  }{\rho_{c}} \frac{n_{X}(t_{now}) }{s(t_{now}) }\, s(t_{now})
=\frac{M_{X} \,s(t_{now})  }{\rho_{c}}   Y_X(t_{now})   \, \,,
\label{eq-OX} 
\eeq
where, defining $H= 100\, h \,{\rm km\,s^{-1}\, Mpc^{-1}}$,
\beq
\rho_c =\frac{3 H^2}{8 \pi G} =1.88 \times 10^{-26} \,h^2 \,{\rm \frac{kg}{m^3}} \,.
\eeq
The entropy now is obtained from eq.\,(\ref{eq-s}) by putting the CMB temperature $T_{CMB}=2.7255$ K \cite{Akrami:2018odb}:
$s(t_{now})= 2891 /{\rm cm^3}$.
Observationally, the cosmological abundance of cold DM has to be $\Omega_c\approx 0.25$.

The effect of accretion is easily included, considering that $N_{X}  Y_{BH}(t_{ev})$ scales as $R_{acc}^{-1/2}$ in the "light" case, 
while as $R_{acc}^{-5/2}$ in the "heavy" case. In both cases, $\Omega_X$ gets suppressed by the effect of accretion.

\subsection{Evaporation during BH domination}

Using eqs.\,(\ref{eq-OX}), (\ref{eq-YXtnow})  and (\ref{eq-YBHBHev}), we have
\beq
\Omega_{X} =  \frac{M_{X} s(t_{now}) }{\rho_{c}}  \frac{1}{\alpha} N_{X}  
  \frac{15} { 4 \pi^3 g_{*,S}(T_{ev}) }  \left(   \frac{  \pi^3  g_*(T_{ev})   }{ 5   }  \right)^{3/4}
 \left( \frac{\mathcal{G} g_{\star, H}  } {10640 \,\pi } \right)^{1/2}
 \left( \frac{M_{Pl} }{M_{BH}} \right)^{5/2} \,.
\eeq

\begin{figure}[t!]
\vskip .0 cm 
\begin{center}
 \includegraphics[width=12.3cm]{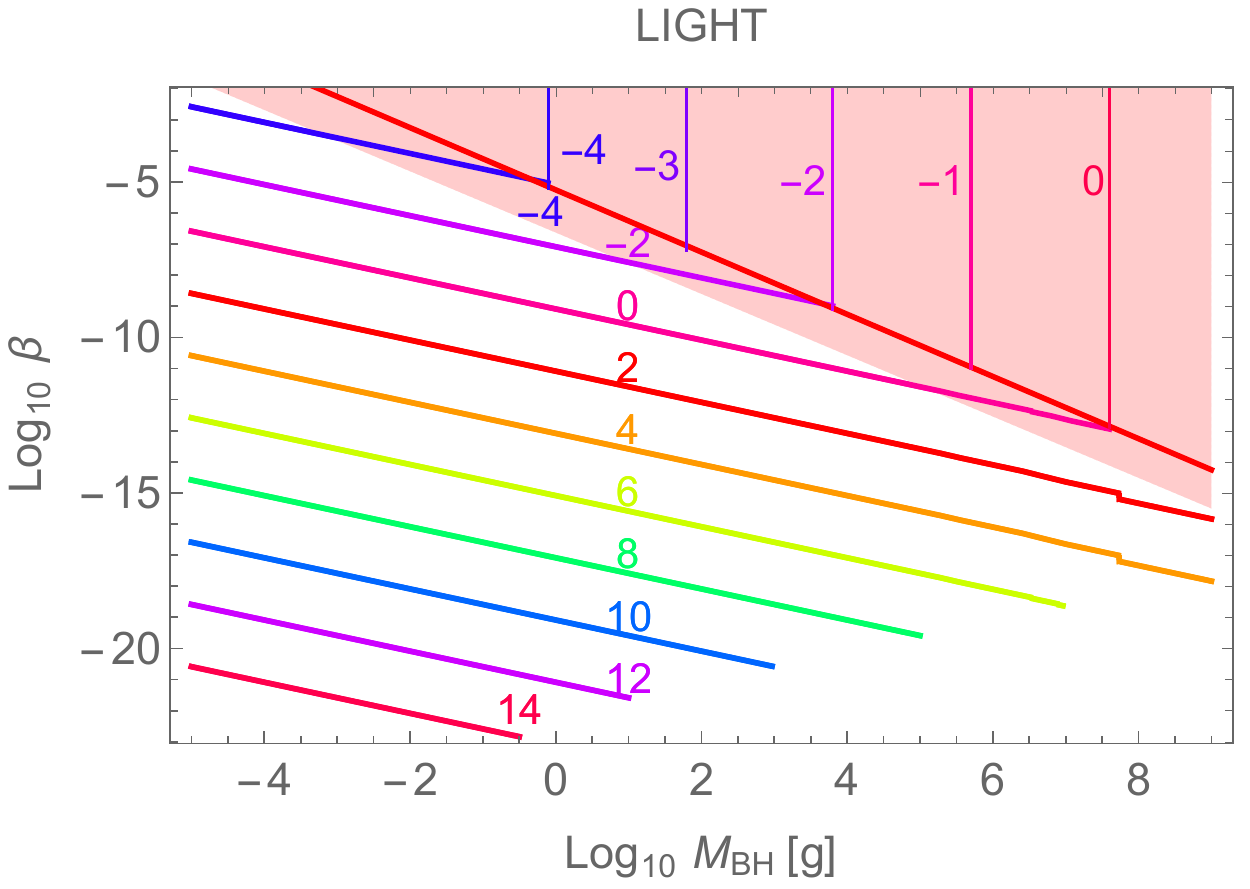}   \\ \vskip1.3cm
  \includegraphics[width=12.3cm]{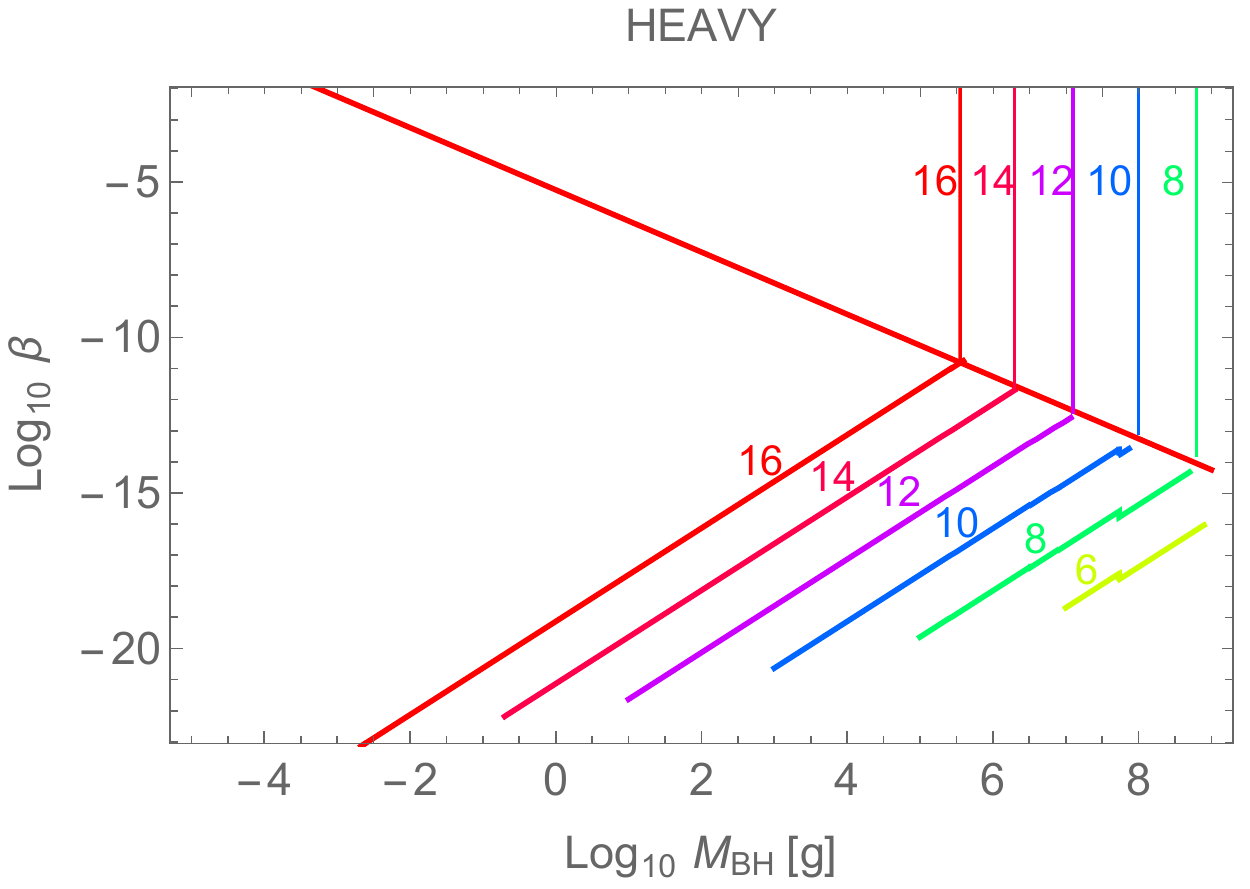}   
 \end{center}
\caption{\baselineskip=15 pt \small  
Values of $\beta$ giving the observed amount of DM today, $\Omega_X \approx 0.25$, in the "light" and "heavy" case respectively. We take $g_{X,H}/g_{*,H}=1/100$, $\alpha=1$, and various values of ${\rm Log}_{10} (M_X c^2[ {\rm GeV}])$, as indicated. }   
\label{fig-DM}
\vskip .0 cm
\end{figure}

In the "light" case, $M_X c^2 < k_B T_{BH}$, we have to plug eq.\,(\ref{eq-NXl}) into the equation above, obtaining
\beq
\Omega_{X} =  \frac{ s(t_{now}) }{\rho_{c}}  \frac{1}{\alpha}    \frac{g_{X,H}}{g_{*,H}} 
  \frac{5} {  \pi^2 g_{*,S}(T_{ev}) }  \left(   \frac{  \pi^3  g_*(T_{ev})   }{ 5   }  \right)^{3/4}
 \left( \frac{\mathcal{G} g_{\star, H}  } {10640 \,\pi } \right)^{1/2}
 \left( \frac{M_{Pl} }{M_{BH}} \right)^{1/2}  M_{X} \,.
 \label{eq-BHl}
\eeq
The BH dominated case is shown in the upper region of the top panel of fig.\,\ref{fig-DM} for the "light" case. Taking $g_{X,H}/g_{*,H}=1/100$ and $\alpha=1$ as reference values, one reproduces the full contribution to DM ($\Omega_X\approx 0.25$), if $M_X c^2 \sim (10^{-4}, 10^{-2},1)$ GeV, for $M_{BH} \sim (1,10^4,10^8)$ g respectively.
Notice that for these values of the $X$ mass, the particles are expected to be too warm today, and fall in the region excluded by the bound on warm DM discussed in sec. \ref{sec-wdm} (see fig. \ref{fig-EDR}), the tension being at the level of a factor of $20$, 
for $g_{X,H}/g_{*,H}=1/100$ and $\alpha=1$. 
For this reason, the shaded (red) region in fig.\,\ref{fig-DM} has to be considered as disfavored. 

An obvious way to avoid the tension with warm DM constraints is to postulate that the X particles contribute only in part to the observed DM abundance, say grossly no more than $5\%$ if $g_{X,H}/g_{*,H}=1/100$.

If one instead postulates that X particles give the dominant contribution to DM, there are two ways to alleviate the tension. 

1) Since $\Omega_X \propto g_{X,H}/ (g_{*,H})^{1/2} M_X$, the first way is to decrease such quantity as much as possible, so that the required value of $M_X$ become accordingly larger. Massive scalars ($s=0$), Weyl and Dirac fermions ($s=1/2$), vectors ($s=1$), all go in the wrong direction with respect to fig.\,\ref{fig-DM}, since for them $g_{X,H}\approx1.8, 2, 4, 1.2$ respectively.
Massive particles with $s=3/2, 2$, go instead in the right direction \cite{Lennon:2017tqq}, since for them $g_{X,H}= 0.56, 0.25$ respectively. 
A significant increase in $g_{*,H}$ would help to alleviate the tension independently of the spin of the $X$ particle. This is the case for supersymmetry realized at low energy, which would provide an effective suppression by a factor by at least $1/\sqrt{3}$.
If the X particle is identified with the gravitino of low energy supersymmetry, the value of $M_X$ required to fully account for DM is
about 3 times the one to be read from fig.\,\ref{fig-DM}; given the present bounds on the velocity of warm DM, this scenario is anyway disfavored. Moreover, the specific case of gravitinos is subject to additional strong constraints from BBN \cite{Khlopov:2004tn, Khlopov:2008qy}.

2) Since $\Omega_X \propto \alpha^{-1} M_X$, the second way, suggested by Fujita et al. \,\cite{Fujita:2014hha}, is to allow for entropy non conservation: with $\alpha >1$, one would need an accordingly larger mass of the X particle to reproduce the observed DM, see eq.\,(\ref{eq-BHl}). 
Given the present constraints on warm DM, fig.\,\ref{fig-EDR} shows that $\alpha \approx \alpha' \gtrsim 5$ would be enough to save the "light" case for BH domination. Note that for ref.\,\cite{Fujita:2014hha} a value as large as $\alpha \sim 10$ was necessary, due to their use of $\delta=6$ as reference value (even adopting the less stringent constraints on WDM \cite{Viel:2005qj} then the one used here\,\cite{Irsic:2017ixq}).
As discussed in Fujita et al.\,\cite{Fujita:2014hha}, a known way to have entropy injection after the evaporation of the BHs is to consider the decay of some matter field\,\footnote{For a general discussion on late time entropy injection (also called re-reheating, since it happens after the first re-heating due to inflation) we refer the interested reader to ref.\,\cite{Nardini:2011hu}.}. 
In particular, they assume that there exists a moduli field and that it dominates the universe after the BHs evaporation: the longer the moduli lives, the more entropy is produced when it decays, and the larger is the value of $\alpha$.
 The decay of the moduli should better happen before BBN, in order to preserve the nice predictions for BBN of the standard cosmological model. 
It would then be reasonable to invoke entropy injection as a solution to the "light" case for BH domination, only for BHs which decays sufficiently before BBN. Note also that $\alpha=\alpha'$ for this entropy injection scenario.

Summarizing, a "little conspiracy" of all the above mentioned effects, might allow to resurrect the case of light DM. In any case, a dedicated study for each DM candidate (to include subleading effects due to the different spins) would be needed to assess with more precision the limits on its mass $M_X$.

The inclusion of accretion does not change the above conclusions. Accretion induces a suppression of the X abundance,
$\Omega_X^{acc}= R_{acc}^{-1/2} \Omega_X$, but also a comparable enhancement of the lower bound on $M_X$.
There is thus no net effect.

In the "heavy" case, $M_X c^2 > k_B T_{BH}$, we have instead to use eq.\,(\ref{eq-NXh})
\beq
\Omega_{X} =  \frac{ s(t_{now}) }{\rho_{c}}   \frac{1}{\alpha} \frac{g_{X,H}}{g_{*,H}}  
  \frac{5} { 64 \pi^4 g_{*,S}(T_{ev}) }  \left(   \frac{  \pi^3  g_*(T_{ev})   }{ 5   }  \right)^{3/4}
 \left( \frac{\mathcal{G} g_{\star, H}  } {10640 \,\pi } \right)^{1/2}
 \left( \frac{M_{Pl} }{M_{BH}} \right)^{5/2}  \frac{M_{Pl}^2}{ M_{X}}  \,.
\eeq
As shown in the lower panel of fig.\,\ref{fig-DM}, for $M_{BH}$ in the range $10^6-10^9$ g,
a stable particle with mass in the range $10^{16}-10^{8}$ GeV would be needed. 
Possible candidates might include, for instance: a right-handed neutrino with Majorana mass, but vanishing Dirac mass term; the lightest supersymmetric particle in the case supersymmetry is realized at high scales. 

Note also that accretion in this case has the following effect:  $\Omega_X^{acc}= R_{acc}^{-5/2} \Omega_X$.

\subsection{Evaporation during radiation domination}

Using eqs.\,(\ref{eq-OX}), (\ref{eq-YXtnow}) and (\ref{eq-YBHRev}), we have
\beq
\Omega_{X} =  \frac{M_{X} s(t_{now}) }{\rho_{c}}   N_{X}  \frac{1}{\alpha \,\alpha_{BH}}
  \beta \frac{ 3 }{4}  
   \left(  \frac{45 \gamma^2  } {16 \pi^3 g_*(T_{f}) }   \right)^{1/4}  \left( \frac{M_{Pl}  }{M_{BH}} \right)^{3/2}   \,   \, .
\eeq

In particular, in the "light" case, $M_X c^2 < k_B T_{BH}$, we have
\beq
\Omega_{X}  = \frac{M_{X}  s(t_{now}) }{\rho_{c}}     \frac{g_{X,H}}{g_{*,H}} \frac{1}{\alpha \,\alpha_{BH}}
   \beta  
 \left(  \frac{45 \pi \gamma^2  } {16  g_*(T_{f}) }   \right)^{1/4}  \left( \frac{M_{BH}  }{M_{Pl}} \right)^{1/2} 
   \,  \,.
\eeq
In the upper panel of fig. \ref{fig-DM}, we show the values of $\beta$ that would provide the present DM density, 
for various values of ${\rm Log}_{10} (M_X c^2[ {\rm GeV}])$, as indicated, and assuming $g_{X,H}/g_{*,H}=1/100$, $\alpha=1$. 
One can see that the region close to BH dominance is in part excluded by the constraints on warm DM. 
Anyway, for $M_{BH} < 10^4$ g, there is an interesting region of parameter space for which GeV scale DM candidates are allowed.
As already noted, accretion has the effect of reducing the $X$ abundance, $\Omega_X^{acc}= R_{acc}^{-1/2} \Omega_X$, hence to increase the $X$ mass.

In the "heavy" case, $M_X c^2 > k_B T_{BH}$,  we have
\beq
\Omega_{X}  = \frac{M_{X} s(t_{now})}{\rho_{c}}     \frac{g_{X,H}}{g_{*,H}}  \frac{1}{\alpha \,\alpha_{BH}}
 \beta 
 \left(  \frac{45  \pi  \gamma^2  } {16  g_*(T_{f}) }   \right)^{1/4} \left(  \frac{M_{Pl}^{7}   }{M_{BH}^{3} M_{X}^{4}} \right)^{1/2}   \,   \,.
\eeq
The bottom panel of fig.\,\ref{fig-DM} shows the values of $\beta$ allowing X to fully reproduce DM. 
In the radiation dominated case, GUT-scale DM particles might be obtained even from BHs as light as 1 g.
In this case, accretion has the following effect:  $\Omega_X^{acc}= R_{acc}^{-5/2} \Omega_X$.

\subsection{BH remnants as dark matter}

\begin{figure}[t!]
\vskip .2 cm 
\begin{center}
 \includegraphics[width=11cm]{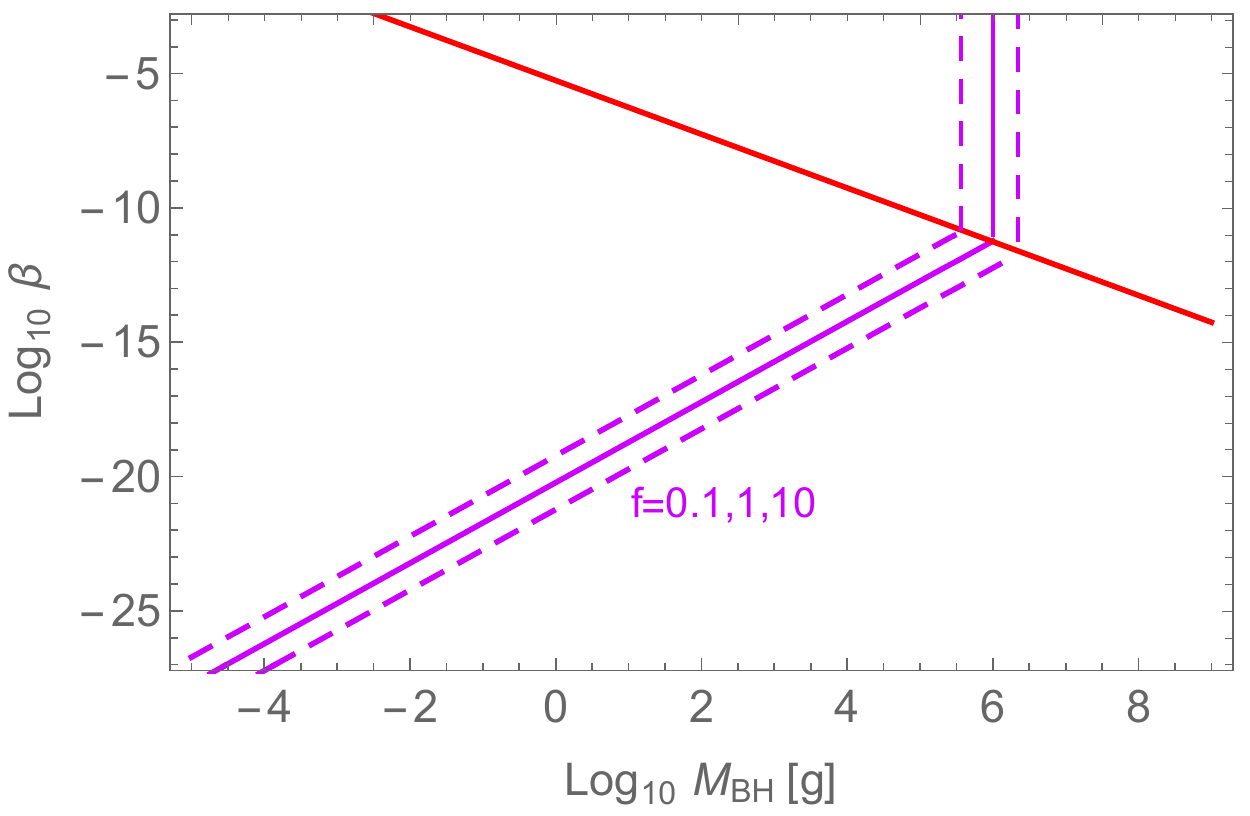}   
 \end{center}
\caption{\baselineskip=15 pt \small  
The values of $\beta$ giving the observed amount of DM as a function of the initial BH mass
in terms of  BH relics with $M_{BH}^R=f M_{Pl}$. From left to right, $f=0.1,1,10$. }   
\label{fig-DMRem}
\vskip .2 cm
\end{figure}

PBH could cease to evaporate when the mass is of order of the Planck mass\,\cite{MacGibbon:1987my},  
$M_{Pl} \approx 2 \times 10^{-5}$\,g. Such relics could constitute a fraction or all of the DM. Their present cosmological abundance is 
\beq
\Omega^R_{BH} =\frac{\rho^R_{BH}   }{\rho_{c}}=\frac{M_{BH}^R \,s(t_{now}) }{\rho_{c}} \frac{n_{BH}(t_{now}) }{s(t_{now}) }\, 
=\frac{M_{BH}^R \,s(t_{now})  }{\rho_{c}}  \frac{1}{\alpha} Y_{BH}(t_{ev})   \, 
\,,
\eeq
For radiation domination one has  to use eq.\,(\ref{eq-YBHRev}), while for BH domination eq.\,(\ref{eq-YBHBHev}).
In the case of BH domination, the calculation was already done by Baumann et al. \cite{Baumann:2007yr}, showing that BH relics could be the DM only for $M_{BH} \sim  10^6$ g.  We find full agreement, as shown in fig.\,\ref{fig-DMRem} for various values of $f=M_{BH}^R/M_{Pl}$. In addition, we extend the calculation to the case of radiation domination and display the required values of $\beta$ as a function of the initial BH mass. For both radiation and BH domination, the effect of accretion is $\Omega_X^{acc}= R_{acc}^{-5/2} \Omega_X$.

\section{Stable particles as dark radiation}

The contribution of a dark energy component to the effective number of relativistic dofs is parametrized by 
\beq
\Delta N_{eff} = \frac{\rho_{DR}(T_{EQ})}{\rho_{R}(T_{EQ})}   \left(  N_\nu + \frac{8}{7} \left(  \frac{11}{4}\right)^{4/3} \right) \,,
\label{eq-defNeff}
\eeq
where $N_\nu=3.045$ \cite{deSalas:2016ztq}, $T_{EQ}$ is the temperature at matter-radiation equality.

Here we review the argument followed in ref.\,\cite{Hooper:2019gtx} to calculate $\Delta N_{eff}$, 
extending it to include a possible entropy non conservation.
The ratio of the energy density in DR with respect to radiation at matter-radiation equality can be rewritten as
a product of three factors
\beq
\frac{\rho_{DR}(T_{EQ})}{\rho_{R}(T_{EQ})} 
=\frac{\rho_{DR}(T_{EQ})}{\rho_{DR}(T_{ev})} \frac{\rho_{DR}(T_{ev})}{\rho_{R}(T_{ev})} 
 \frac{\rho_{R}(T_{ev})}{\rho_{R}(T_{EQ})} \,.
\eeq

The first factor is the dilution of the DR energy density as the universe expands and cool
\beq
\frac{\rho_{DR}(T_{EQ})}{\rho_{DR}(T_{ev})}=\left( \frac{a_{ev}}{a_{EQ}}   \right)^4 \,\,.
\label{eq-fDREQev}
\eeq

The third factor can be calculated as follows. 
As the universe expands and cool, the energy density in  radiation is diluted additionally by a series of transfers. As done in the previous sections, we allow for entropy non conservation, $\alpha' (a^3 s)_{ev}= (a^3 s)_{EQ}$, so that
\beq
{\alpha'} a_{ev}^3 \, g_{*,S} (T_{ev})\,  T^3_{ev} = a_{EQ}^3 \, g_{*,S} (T_{EQ})\,  T^3_{EQ}\,.
\eeq
Recall that $g_{*,S}=g_{*}$ at high temperatures, but at matter-radiation equality, when $k_B T_{EQ} =0.75$ eV,
we have $g_{*,S} (T_{EQ})=3.94$, while $g_{*} (T_{EQ})=3.38$.
Thus
\beq
\frac{T_{EQ}}{T_{ev}}  = (\alpha')^{1/3} \,\frac{a_{ev}}{a_{EQ}}  \left(  \frac{g_{*,S} (T_{ev})  }{g_{*,S} (T_{EQ})} \right)^{1/3}\,,
\label{eq-TEQev}
\eeq
so that 
\beq
\frac{\rho_{R}(T_{EQ})}{\rho_{R}(T_{ev})}= \frac{g_*(T_{EQ})}{g_*(T_{ev})} \left(  \frac{T_{EQ}}{T_{ev}} \right)^4
= \frac{g_*(T_{EQ})}{g_*(T_{ev})}  {\alpha'}^{4/3} \left( \frac{a_{ev}}{a_{EQ}} \right)^4  
\left(  \frac{g_{*,S} (T_{ev})  }{g_{*,S} (T_{EQ})} \right)^{4/3}\,.
\eeq
Since at $T_{ev}$ we have $g_{*,S}=g_{*}$, this can be simplified by
\beq
\frac{\rho_{R}(T_{EQ})}{\rho_{R}(T_{ev})}
= {\alpha'}^{4/3}  \left( \frac{a_{ev}}{a_{EQ}} \right)^4      \frac{g_*(T_{EQ})   }{   g_{*,S} (T_{EQ}) }  \frac{  g_{*,S} (T_{ev}) ^{1/3}  } {   g_{*,S} (T_{EQ})^{1/3} } \,.
\label{eq-fREQev}
\eeq

Using eqs.\,(\ref{eq-fDREQev}) and (\ref{eq-fREQev}), the ratio of the energy density in DR with respect to radiation at matter-radiation equality is
\beq
\frac{\rho_{DR}(T_{EQ})}{\rho_{R}(T_{EQ})} 
=\frac{1} {{\alpha'}^{4/3}} \, \frac{\rho_{DR}(T_{ev})}{\rho_{R}(T_{ev})}\,  \frac{   g_{*,S} (T_{EQ}) } {g_*(T_{EQ})   }  \frac{   g_{*,S} (T_{EQ})^{1/3} } {  g_{*,S} (T_{ev}) ^{1/3}  }\,.
\eeq

Substituting the above expression in eq.\,(\ref{eq-defNeff}), one has 
\beq
\Delta N_{eff} =  \frac{1}{{\alpha'}^{4/3} }\,  \frac{\rho_{DR}(T_{ev})}{\rho_{R}(T_{ev})}  
  \frac{   g_{*,S} (T_{EQ}) } {g_*(T_{EQ})   } 
 \frac{   g_{*,S} (T_{EQ})^{1/3} } {  g_{*,S} (T_{ev}) ^{1/3}  } 
 \left(  N_\nu + \frac{8}{7} \left(  \frac{11}{4}\right)^{4/3} \right) 
\approx 2.9\,\frac{1}{ {\alpha'}^{4/3}} \, \frac{\rho_{DR}(T_{ev})}{\rho_{R}(T_{ev})}  \,,
\eeq
in agreement with \cite{Hooper:2019gtx} for $\alpha'=1$.

\subsection{BH domination}

Consider first the case of BH domination. After evaporation, the fraction of the universe energy density in DR in such particles is simply given be the proportion of their dofs \cite{Hooper:2019gtx}: 
\beq
\frac{\rho_{DR}(T_{ev})}{\rho_{R}(T_{ev})}=\frac{g_{DR,H}}{g_{*,H}} \,\,.
\eeq
Numerically, one has
\beq
\Delta N_{eff} \approx  2.9 \, \,\frac{1}{ {\alpha'}^{4/3}} \, \frac{g_{DR,H}}{g_{*,H}}\,.
\eeq

\begin{figure}[t!]
\vskip .2 cm 
\begin{center}
 \includegraphics[width= 11.2cm]{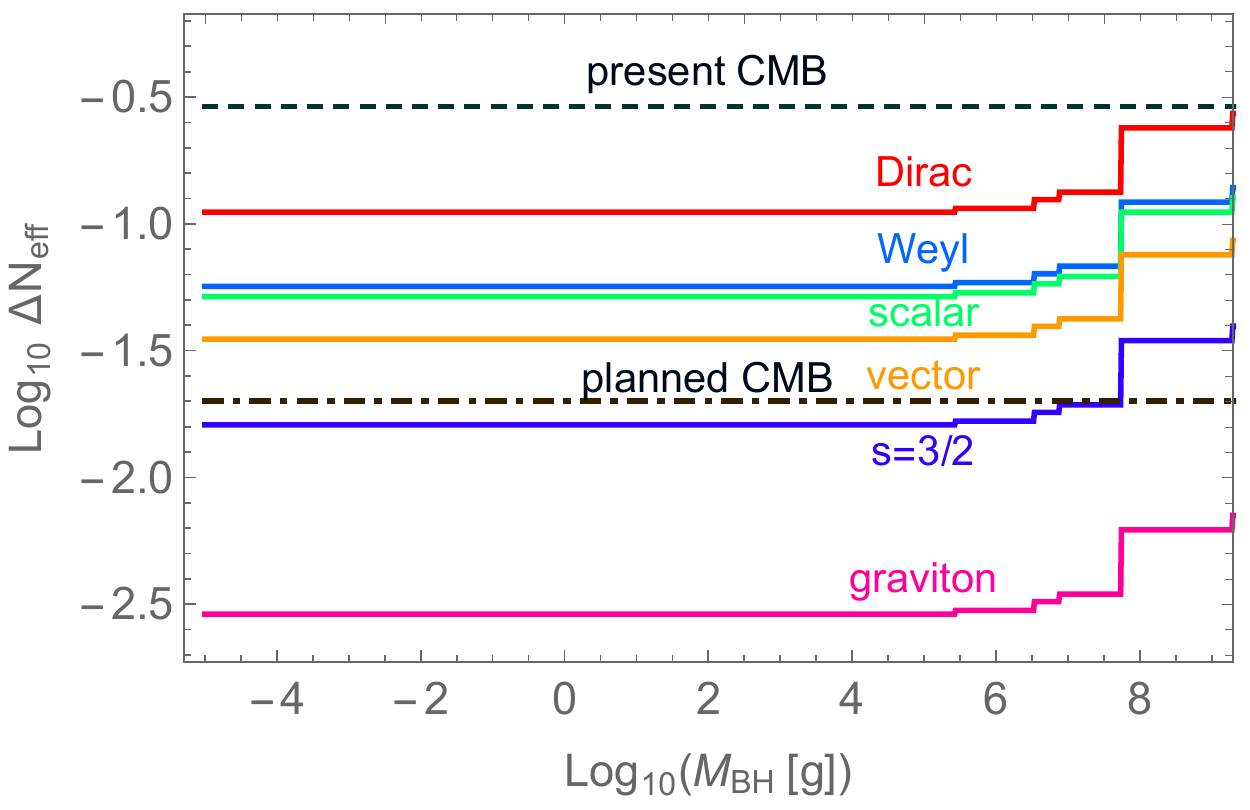}  \,\,\, 
 \end{center}
\caption{\baselineskip=15 pt \small  
Various DR contributions to $\Delta N_{eff}$, as a function of the BH mass, assuming an epoch of BH domination and taking $\alpha'=1$.}   
\label{fig-DR}
\vskip .4 cm
\end{figure}

To be more precise, one should include the implicit dependence of $g_{*,S}(T_{ev})$ on $M_{BH}$, as shown in fig.\,\ref{fig-DR}.
For $M_{BH}< 6\times 10^8$ g, all the SM has to be included, otherwise only a part of it. 
Fig. \ref{fig-DR} shows the values of $\Delta N_{eff}$ for various particles, taking $\alpha'=1$. 
From top to bottom, we consider: a Dirac and a Weyl fermion ($g_{DR,H}=4$ and $2$ respectively), a scalar ($g_{DR,H}=1.82$), a massive vector $g_{DR,H}=1.23$, a massive $s=3/2$ particle ($g_{DR,H}=0.56$), and a massless graviton ($g_{DR,H}=0.1$).
These results are in full agreement with \cite{Hooper:2019gtx}.
The present sensitivity to $\Delta N_{eff}$ of CMB observations is also shown: since $N_{eff} = 2.99 \pm 0.17$\,\cite{Aghanim:2018eyx},
one has $ N_{eff}<3.33$ at $2\sigma$, or equivalently $\Delta N_{eff} =N_{eff}-N_\nu  <  0.29$ at $2\sigma$.
Interestingly enough, there are optimistic possibilities of detecting some signal in the future \cite{Hooper:2019gtx},
as the predicted contribution to $\Delta N_{eff}$ is potentially within the projected reach of stage IV experiments, $\Delta N_{eff} \approx 0.02$: this is the case for DR particles with $s < 3/2$.

Let us focus on some specific candidate. A scalar, such as the axion or the majoron, would give $\Delta N_{eff} \approx  0.052$. It would be possible to have $N_a$ axions; at present, 6 axions would already exceed the present limit, $\Delta N_{eff} <  0.29$ at $2\sigma$.
Among Weyl fermion candidates, one could consider the possibility of stable right-handed (sterile) neutrinos. For example, within a seesaw realized at low energies (the left-handed neutrinos are already included in the SM), one has to consider the contribution of at most one stable right-handed neutrino\footnote{At least two right-handed neutrinos have to provide neutrino masses and mixing: they would be short lived because of the decays mediated by the Dirac couplings.}, which would give $\Delta N_{eff}\approx 0.056$. 
In general, one could extend the seesaw to include many additional sterile neutrinos: at present, 6 sterile neutrinos would provide a contribution exceeding the present limit on $\Delta N_{eff}$.

The effect of entropy non conservation is to suppress $\Delta N_{eff}$. Notice that the amount of entropy non conservation that would save the light DM scenario with BH domination, $\alpha' \sim 5$, would imply a suppression of $\Delta N_{eff}$ by about the same factor. Even in this case, the DR contribution to $\Delta N_{eff}$ might be at hand of future detection for $s<2$.

As already discussed, low energy (about 10 TeV) supersymmetric models imply an enhancement in $g_{*,H}$ by at least a factor of 3, for BHs with masses below $10^8$ g. In this case, the contribution to $\Delta N_{eff}$ is accordingly suppressed by a factor of 3. 
There would be no chance to detect DR for BHs with masses below $10^8$ g: a scalar and a Weyl fermion would indeed give respectively $\Delta N_{eff} \approx  0.017$ and $\Delta N_{eff} \approx  0.019$.

Notice also that  accretion has no effect on $\Delta N_{eff}$ for BH domination.

\subsection{Radiation domination}

Consider now the case of radiation domination. After evaporation, the fraction of the universe energy density in DR in such particles is 
\beq
\frac{\rho_{DR}(T_{ev})}{\rho_{R}(T_{ev})}
=\frac{g_{DR,H}}{g_{*,H}} \frac{\rho_{BH}(t_{ev})}{\rho_{R}(t_{ev})}
=\frac{g_{DR,H}}{g_{*,H}} f(t_{ev})\,.
\eeq
Inserting eqs. (\ref{eq-ftevftf}) and (\ref{eq-beta}), in the equation above, one obtains
\beq
\frac{\rho_{DR}(T_{ev})}{\rho_{R}(T_{ev})}
=\frac{g_{DR,H}}{g_{*,H}}  \frac{\beta}{\bar \beta}   \,.
\eeq
Explicitly, the radiation dominated case is suppressed with respect to the BH dominated case by 
\beq
\frac{\Delta N_{eff}^R}{ \Delta N_{eff}} =  \frac{\beta}{\bar \beta} = \beta 
 \gamma^{1/2}  \left( \frac{10640 \,\pi}{{\mathcal G} \, g_{*,H}(T_{BH})} \right)^{1/2}  \frac{M_{BH}}{ M_{Pl}} \,.
\eeq

Even in the case of radiation domination, for the highest possible values of $\beta$ (those close to the BH domination region),
there would be the possibility to detect some signal.
While accretion has no effect for BH domination, in the radiation dominated case, $ {\Delta N_{eff}^R} \propto R_{acc}^{3/2}$.

\section{Discussion and conclusions}
\label{sec-concl}

We have reconsidered the issue of generating DM and DR via the mechanism of evaporation of PBHs, including explicitly the accretion effect and the impact of a possible non conservation of entropy. We considered both cases of BH and radiation domination. The first one is particularly appealing since the DM abundance and the contribution to $\Delta N_{eff}$ do not depend on $\beta$, the fraction of the energy density in BHs over radiation at the time of PBHs formation.

The possible stable candidates for DM from PBHs grossly divide in two categories: either they are very heavy, see fig.\ref{fig-DM},
or very light, in the MeV-GeV range. Another interesting possibility for DM are Planck scale remnants. 
In our opinion, very heavy stable particles might not be easy to justify theoretically, and the scenario of light stable particles would be more attractive. However, it turns out that the MeV-GeV stable DM candidates required in the scenario of BH domination, and also in a small portion of parameter space for radiation domination, are disfavored: they are produced with so large momenta with respect to their mass, that they end up being too warm DM candidates now, in conflict with observations on structure formation. 

Given the elegance of the BH domination scenario, it is important to understand how robustly the light DM case is excluded. DM particles with a high value of the spin (as for instance gravitinos) go in the direction of alleviating the tension\,\cite{Lennon:2017tqq}. A significant increase in $g_{*,H}$, as would be the case for low energy supersymmetry, would also help. Some amount of entropy non conservation\,\cite{Nardini:2011hu, Fujita:2014hha}, at the level of $\alpha \sim 5$, seems enough to resurrect the light DM scenario\,\footnote{In ref.\,\cite{Fujita:2014hha} a value as large as $\alpha \sim 10$ was necessary, due to their use of $\delta=6$ as reference value.}. The effect of accretion would instead have no impact. In our opinion, an ad hoc analysis for each specific DM candidate would be necessary to assess more robustly the issue of the tension with structure formation. 

As for DR, we confirm the interesting results of ref.\cite{Hooper:2019gtx}, that future observations might be sensitive to the DR contribution of light stable particles emitted by PBHs, especially those with lower values of the spin, see fig.\,\ref{fig-DR}. 
This applies to the case of BH domination and, in part, also to the case of radiation domination.  
The amount of entropy non conservation that would rescue the light DM scenario ($\alpha \sim 5$), 
would suppress the DR contribution to $\Delta N_{eff}$ down to values that might be anyway at hand of the planned experimental sensitivity for $s<2$.

There are many candidates for stable particles which could be detected via their contribution to $\Delta N_{eff}$, in the case of BH domination. For instance, in the category of scalars, axions and majorons are interesting candidates. For $s=1/2$ fermions, one could consider: supersymmetric particles such as neutralinos, axinos; very light right-handed (sterile) neutrinos, as those of the $\nu$MSM \cite{Asaka:2005an, Asaka:2005pn}; Dirac light neutrinos \,\cite{Lunardini:2019zob}.

Let us focus for instance on one specific example as a concrete application of our general results.
It is easy to reconsider the $\nu$MSM \cite{Asaka:2005an, Asaka:2005pn} in the light of the scenario of particle production from evaporation from PBHs: two GeV-scale right-handed neutrinos, together with two left-handed neutrinos, would realize a low energy seesaw mechanism, fully explaining the phenomenology of neutrino masses and mixings. They would be short lived because of the large mixing in the Dirac mass term, and might explain the baryon asymmetry via leptogenesis. 
With small enough Dirac coupling, the third right-handed neutrino could instead be long lived on cosmological 
times. It would dominantly decay in three left-handed neutrinos, but the subdominant decay mode in a left-handed neutrino and a photon would be very interesting: the monochromatic photon emitted having just energy equal to half of the right-handed neutrino mass. Such decay mode could be used as an independent check of the scenario. A stable right-handed neutrino with MeV-GeV mass, would contribute to DM. From the discussion above, in the case of BH domination it would be too warm, unless invoking entropy non conservation; in the case of radiation domination, it might have a GeV scale mass, for BHs lighter than about $10^4$ g. 
Giving up the role of the third right-handed neutrino as DM, one could instead explore its role as DR.
If sufficiently light, say grossly below the keV, the third right-handed neutrino would indeed significantly contribute to DR: in the case of BH domination its contribution to $\Delta N_{eff}$ would be at hand of the future experimental sensitivity. 
In addition, in the case that the $3.5$ keV line is generated by the right-handed neutrino subdominant decay, its mass would be 
$7$ keV: this value would require $M_{BH}\gtrsim 10^2$ g, as can be seen from fig.\,\ref{fig-EDR}. 

To conclude, the DM and DR production mechanism from the process of PBHs evaporation is an interesting and open issue, 
also in view of its connection with gravitational waves\,\cite{Fujita:2014hha, Inomata:2020lmk, Hooper:2020evu}. 
From the point of view of model building, it is a fascinating arena where to study different DM and DR candidates beyond the SM.

\section*{\large Acknowledgements}

We thank the CERN Theory Department for kind hospitality and support during the completion of this work.
We acknowledge partial support by the research project TAsP (Theoretical Astroparticle Physics) funded by the Istituto Nazionale di Fisica Nucleare (INFN). We thank M. Viel for useful discussions and the anonymous referee for his valuable suggestions to improve the paper.

\bibliographystyle{elsarticle-num} 
\bibliography{bib} 
\end{document}